\documentclass[a4paper,UKenglish,cleveref, autoref]{lipics-v2021}

\usepackage{graphicx} 
\usepackage{stmaryrd}
\usepackage{amsmath}
\usepackage{amssymb}
\usepackage{amsthm}
\usepackage{hyperref}

\usepackage{xspace}
\usepackage{todonotes}
\usepackage[ruled,noend,linesnumbered]{algorithm2e}
\usepackage[capitalise]{cleveref}
\usetikzlibrary{positioning,shapes,shapes.misc,fit,positioning,calc}
\usepackage{subcaption}
\usepackage[bibliography=common,appendix=inline]{apxproof}
\newtheoremrep{lemma}{Lemma}
\newtheoremrep{theorem}{Theorem}

\newcommand{\N}{\ensuremath{\mathbb{N}}}
\newcommand{\Z}{\ensuremath{\mathbb{Z}}}
\newcommand{\R}{\ensuremath{\mathbb{R}}}
\newcommand{\xto}{\ensuremath{\xrightarrow}}

\newcommand{\veco}{\ensuremath\overrightarrow 0}
\newcommand{\nodelay}{\ensuremath{}\smallsetminus\!\!\!\!\!\nearrow}

\makeatletter
\newsavebox{\@brx}
\newcommand{\llangle}[1][]{\savebox{\@brx}{\(\m@th{#1\langle}\)}%
  \mathopen{\copy\@brx\kern-0.5\wd\@brx\usebox{\@brx}}}
\newcommand{\rrangle}[1][]{\savebox{\@brx}{\(\m@th{#1\rangle}\)}%
  \mathclose{\copy\@brx\kern-0.5\wd\@brx\usebox{\@brx}}}
\makeatother

\newcommand{\coal}[1]{\ensuremath\llangle#1\rrangle}
\newcommand{\coalbr}[1]{\ensuremath\llbracket#1\rrbracket}

\newcommand{\until}{\ensuremath\mathcal{U}}

\newcommand{\uppaal}{\textsc{Uppaal}\xspace}
\newcommand{\uppaaltiga}{\textsc{Uppaal Tiga}\xspace}
\newcommand{\tiga}{\textsc{Tiga}\xspace}

\newcommand{\ExamColIRaw}{red!80!black}
\newcommand{\ExamColIIRaw}{green!80!black}
\newcommand{\ExamColIIIRaw}{blue!70!black}
\newcommand{\ExamColI}{\color{\ExamColIRaw}}
\newcommand{\ExamColII}{\color{\ExamColIIRaw}}
\newcommand{\ExamColIII}{\color{\ExamColIIIRaw}}
\newcommand{\ExamPlayerI}{\textrm{\ExamColI I}\xspace}
\newcommand{\ExamPlayerII}{\textrm{\ExamColII II}\xspace}
\newcommand{\ExamPlayerIII}{\textrm{\ExamColIII III}\xspace}

\newcommand*{\bhighlight}{\tikz[remember picture,overlay]{\node (highlight_begin) {};}\ignorespaces}
\newcommand*{\ehighlight}[1]{\tikz[remember picture,overlay]{
    \node (origo) {};
    \node[fill=#1,opacity=.24,fit={(highlight_begin.north)($(current page.west |- origo)+(0.66\paperwidth-0.5\textwidth+1mm,0)$)($(current page.east |- origo)+(-0.38\paperwidth+0.5\textwidth-1mm,0)$)}] {};}\ignorespaces
}

\hideLIPIcs
\nolinenumbers

\makeatletter
\renewcommand*{\@fnsymbol}[1]{\ensuremath{\ifcase#1\or \dagger\or \ddagger\or
   \mathsection\or \mathparagraph\or \|\or **\or \dagger\dagger
   \or \ddagger\ddagger \else\@ctrerr\fi}}
\makeatother

\title{On-The-Fly Symbolic Algorithm for Timed ATL with Abstractions}

\author{Nicolaj \O. Jensen}{Department of Computer Science, Aalborg University, Aalborg 9220, Denmark}{noje@cs.aau.dk}{https://orcid.org/0009-0005-2359-204X}{}

\author{Kim G. Larsen}{Department of Computer Science, Aalborg University, Aalborg 9220, Denmark}{kgl@cs.aau.dk}{https://orcid.org/0000-0002-5953-3384}{}

\author{Didier Lime}{Nantes Université, École Centrale Nantes, CNRS, LS2N, UMR 6004, F-44000 Nantes, France}{Didier.Lime@ec-nantes.fr}{https://orcid.org/0000-0001-9429-7586}{}

\author{Ji\v{r}\'{i} Srba}{Department of Computer Science, Aalborg University, Aalborg 9220, Denmark}{srba@cs.aau.dk}{https://orcid.org/0000-0001-5551-6547}{}

\authorrunning{N.\,\O. Jensen, K.\,G. Larsen, D. Lime, and J. Srba}



\Copyright{Nicolaj \O. Jensen, Kim G. Larsen, Didier Lime, and Ji\v{r}\'{i} Srba} 

\supplementdetails[subcategory={Reprod.\ Package}]{Software}{https://doi.org/10.5281/zenodo.15195408}

\funding{VILLUM INVESTIGATOR project S4OS and ANR project BisoUS ANR-22-CE48-0012.}

\begin{document}

\maketitle

\begin{abstract}
    Verification of real-time systems with multiple components controlled by multiple parties is a challenging task due to its computational complexity.
We present an on-the-fly algorithm for verifying timed alternating-time temporal logic (TATL), a branching-time logic with quantifiers over outcomes that results from coalitions of players in such systems.
We combine existing work on games and timed CTL verification in the abstract dependency graph (ADG) framework, which allows for easy creation of on-the-fly algorithms that only explore the state space as needed.
In addition, we generalize the conventional inclusion check to the ADG framework which enables dynamic reductions of the dependency graph.
Using the insights from the generalization, we present a novel abstraction that eliminates the need for inclusion checking altogether in our domain.
We implement our algorithms in \uppaal and our experiments show that while inclusion checking considerably enhances performance, our abstraction provides even more significant improvements, almost two orders of magnitude faster than the naive method.
In addition, we outperform \uppaaltiga, which can verify only a strict subset of TATL.
After implementing our new abstraction in \uppaaltiga, we also improve its performance by almost an order of magnitude.

\end{abstract}

\section{Introduction}

\begin{figure}
    \centering
    \begin{subfigure}{\textwidth}
    \centering 
        \resizebox{.8\textwidth}{!}{%
            \begin{tikzpicture}[node distance=1.65,
                        loc/.style = {draw,fill=lightgray!50,circle,text width=3.5mm, align=center},
                        arr/.style = {-stealth}
                    ]
                \node[loc,label=above left:$\mathbf{A,x\leq 4}$] (A) {};
                \node[loc,label=above left:$\mathbf{B,x\leq 5}$, right=of A] (B) {};
                \node[loc,label=above left:$\mathbf{C}$, above right=of B] (C) {};
                \node[loc,label=below right:$\mathbf{D,x\leq 3}$, below right=of B] (D) {};
                \node[loc,label=below right:$\mathbf{Goal}$, below right=of C] (Goal) {};
        
                \draw[arr] (A.west) ++ (-0.4,0) -- (A);
                \draw[arr,\ExamColIRaw] (A) -- node {\ExamPlayerI} node[below=1.2mm,black] {$a_1$} (B);
                \draw[arr,\ExamColIRaw] (B) -- node {\ExamPlayerI} node[below=2mm,black] {$a_2$} (C);
                \draw[arr,\ExamColIIIRaw] (B) -- node {\ExamPlayerIII} node[below=1.2mm,black] {$a_3,x\leq 2$} (Goal);
                \draw[arr,\ExamColIIRaw] (B) -- node {\ExamPlayerII} node[below left,black] {$a_4,x\leq 3$} (D);
                \draw[arr,\ExamColIIRaw] (C) -- node {\ExamPlayerII} node[below=2mm,black] {$a_5$} (Goal);
                \draw[arr,\ExamColIIIRaw] (D) -- node {\ExamPlayerIII} node[below=2mm,black] {$a_6$} (Goal);
        
                \node[right=of Goal,text width=55mm] {
                    $\mathcal A\nvDash\coal{\ExamPlayerI}\Diamond\mathbf{Goal}$\\
                    $\mathcal A\vDash\coal{\ExamPlayerII}\Diamond\mathbf{Goal}$\\
                    $\mathcal A\nvDash\coal{\ExamPlayerIII}\Diamond\mathbf{Goal}$\\
                    $\mathcal A\vDash\coal{\ExamPlayerI, \ExamPlayerIII}\Diamond\mathbf{Goal}$\\
                    $\mathcal A\vDash\coalbr{\ExamPlayerII}\Diamond\mathbf{Goal}$\\
                    $\mathcal A\nvDash\coal{\ExamPlayerII}(\neg \mathbf{C}\;\until\;\mathbf{Goal})$\\
                    $\mathcal A\vDash\coal{\ExamPlayerI, \ExamPlayerIII}(\neg\mathbf{C}\;\until\;\mathbf{Goal})$\\
                    $\mathcal A\vDash\coal{\ExamPlayerI}\Box\neg\coal{\ExamPlayerIII}\Diamond\mathbf{Goal}$\\
                    $\mathcal A\nvDash\coal{\ExamPlayerII}\Box\neg\coal{\ExamPlayerIII}\Diamond\mathbf{Goal}$\\
                    $\mathcal A\nvDash\coal{\ExamPlayerII}\Diamond_{< 5}\mathbf{Goal}$\\
                };
                
            \end{tikzpicture}}
        \caption{An example timed game $\mathcal A$ with three players: \ExamPlayerI, \ExamPlayerII, and \ExamPlayerIII and a single clock $x$ that increases as time passes. Multiple TATL properties of the system are shown on the right.}
        \label{fig:intro_example}
    \end{subfigure}
    \begin{subfigure}{\textwidth}
        \small
        $$
                \ExamPlayerI:\langle\ell,\nu\rangle\mapsto\begin{cases}
                    a_1 & \text{if }\ell=\mathbf{A}\\
                    a_2 & \text{if }\ell=\mathbf{B}\land\nu(x)=5\\
                    \lambda & \text{otherwise}
                \end{cases}
                \hspace{6mm} \;
                \ExamPlayerIII:\langle\ell,\nu\rangle\mapsto\begin{cases}
                    a_3 & \text{if }\ell=\mathbf{B}\land\nu(x)\leq 2 \\
                    a_6 & \text{if }\ell=\mathbf{D} \\
                \lambda & \text{otherwise}
            \end{cases}
        $$
        \caption{An example strategy profile for player \ExamPlayerI and \ExamPlayerIII that witnesses $\mathcal A\vDash\coal{\ExamPlayerI,\ExamPlayerIII}(\neg\mathbf{C}\;\until\;\mathbf{Goal})$.}
        \label{fig:intro_example:strategy_profile}
    \end{subfigure}
    \caption{\;}
\end{figure}

\noindent Correctness is essential for many real-time systems, including distributed communication networks, energy grid management, and air traffic control. Ensuring the correctness of such systems is challenging due to the number of concurrent internal and external components. The interleaved behavior of all these components leads to an exponential growth in the number of states, a phenomenon known as the state-space explosion problem~\cite{Clarke2018Handbook}. Addressing this challenge is a core focus of the field of model checking. In this paper, we consider real-time systems with components controlled by more than two parties, some cooperative, others adversarial. Real examples of such systems include electronic circuits with multiple components, perhaps some counterfeit~\cite{Tehranipoor2015CounterfeitICs}, and networks of routers located in different countries~\cite{Edmundson2018NationStateInternet,Anwar2015InvestigatingWildRouting}. We model these systems as timed multiplayer games (TMG) and consider properties described in timed alternating-time temporal logic (TATL) \cite{Clarke2018Handbook,Rajeev2002AtlLogics}.
TMGs are generalized timed game automata~\cite{Maler1995SynthesisTimedSystem,Asarin1998SynthesisTimedAutomata} and can have more than two players. Like in timed games, each discrete action belongs to a specific player and can be executed only at the discretion of that player.
Consider the example TMG in \cref{fig:intro_example} with three players, \ExamPlayerI, \ExamPlayerII, and \ExamPlayerIII, and a single real-valued clock $x$, initially at 0, that increases as time passes. From the starting location, $\mathbf{A}$, player \ExamPlayerIII has no strategy to ensure that the system reaches the $\mathbf{Goal}$ location, expressed as $\mathcal A\nvDash\coal{\ExamPlayerIII}\Diamond\mathbf{Goal}$ in TATL. This is because player \ExamPlayerI can wait in location $\mathbf{A}$ until $x>3$ and then direct the system to location $\mathbf{C}$ without giving \ExamPlayerIII any influence. Here, the system can remain indefinitely. On the other hand, \ExamPlayerII can ensure $\mathbf{Goal}$ is reached as \ExamPlayerI and \ExamPlayerIII cannot leave the system in a location where \ExamPlayerII cannot make progress. TATL also allows us to consider coalitions of collaborating players. If players \ExamPlayerI and \ExamPlayerIII work together, then they can ensure $\mathbf{Goal}$ is reached without visiting location $\mathbf{C}$, expressed as $\mathcal A\vDash\coal{\ExamPlayerI,\ExamPlayerIII}(\neg\mathbf{C}\;\until\;\mathbf{Goal})$. We also consider nested properties, such as $\mathcal A\vDash\coal{\ExamPlayerI}\Box\neg\coal{\ExamPlayerIII}\Diamond\mathbf{Goal}$ which states that \ExamPlayerI can ensure that in all reached states, \ExamPlayerIII does not have a strategy to ensure $\mathbf{Goal}$ is reached. Furthermore, timed properties add conditions on the time passed. For example, $\mathcal A\nvDash\coal{\ExamPlayerII}\Diamond_{< 5}\mathbf{Goal}$ states that \ExamPlayerII cannot guarantee $\mathbf{Goal}$ is reached strictly within 5 time units.

For verification, it is often unnecessary to explore the entire state space to determine the correctness of a property. One approach that reduces unnecessary work is to decompose the problem into sub-problems and lazily structure them in a dependency graph~\cite{LiuSmolka1998LinFixedPoint,Enevoldsen2019ADGs,Enevoldsen2022EADGs}. As the process unfolds, sub-problems eventually become trivial, and their solutions are used to resolve the more complex dependent problems. In many cases, the root problem can be answered without first constructing the entire dependency graph and the process can terminate early. This framework was formalized in~\cite{Enevoldsen2019ADGs,Enevoldsen2022EADGs} using extended abstract dependency graphs (EADGs) and has been used to create on-the-fly algorithms for other model checking domains~\cite{Carlsen2023CGAAL,Cassez2005TimedGames,Jensen2016WeightedCTL}.

\paragraph*{Our contributions}
We develop an on-the-fly TATL algorithm within the EADG framework by providing a sound encoding of the problem. Our approach builds upon previous work involving encodings for timed systems and alternating-time temporal logic, showing their orthogonality, and we provide many semantic details that were left out in previous work.
Since TATL is a~superset of TCTL~\cite{Alur1990MCRealTime,Alur1993DenseRealTime,Rajeev2002AtlLogics}, our algorithm is also the first on-the-fly algorithm for TCTL with the generic freeze operator.
Moreover, we formalize vertex merging in EADGs as a~generalization of inclusion checking from prior work~\cite{Cassez2005TimedGames} and incorporate it into our encoding.
In addition, this generalization also leads to an expansion abstraction that removes the need for conventional inclusion checking in our encoding altogether.
Finally, we implement and evaluate various configurations of our algorithm and compare their performance to the state-of-the-art model checker for real-time games~\uppaaltiga~\cite{Cassez2005TimedGames,Behrmann2007Tiga}. Our configuration with inclusion checking is, as expected, on par with \tiga on the subset of formulae that \tiga can handle, however, incorporating our expansion abstraction improves performance by nearly an order of magnitude.

\paragraph*{Related work}
The tool \uppaaltiga~\cite{Cassez2005TimedGames,Behrmann2007Tiga}, which has since been integrated in \uppaal~\cite{Hendriks2006Uppaal}, uses a similar method to solve a simpler version of the problem: strategy synthesis in two-player timed games. 
In fact,~\cite{Cassez2005TimedGames} was published 20 years ago at CONCUR'05, and our paper brings the foundational work of that paper into the modern framework and expands it to the broader TATL logic. \tiga was the first to adapt the dependency graph-based on-the-fly algorithm by Liu and Smolka~\cite{LiuSmolka1998LinFixedPoint} to model checking, and the idea was eventually generalized into the EADG framework~\cite{Enevoldsen2019ADGs,Enevoldsen2022EADGs}. \tiga relies on inclusion checking to merge vertices and improve performance, a detail that has not been generalized to the EADG framework until this paper. This paper also includes many semantic details that were left out in \cite{Cassez2005TimedGames}.
Alternating-time temporal logic (ATL), defined by T.~Henzinger and R.~Alur in \cite{Rajeev2002AtlLogics} as an extension to branching-time logic~\cite{Clarke2018Handbook}, introduces ways to quantify over possible outcomes resulting from coalitions of players working together. ATL properties with a single coalition can be reduced to a synthesis problem in a two-player game, a problem that has already been intensively studied~\cite{Maler1995SynthesisTimedSystem,Asarin1998SynthesisTimedAutomata,deAlfaro2001SymbolicAlgoInfiniteGames,Cassez2005TimedGames}. However, it is also possible to nest the coalition quantifiers to express more intriguing properties. Recently, Carlsen et~al.~\cite{Carlsen2023CGAAL} used the EADG framework for on-the-fly verification of (untimed) ATL properties in concurrent games with multiple players. We extend this work to the setting of timed games and timed logics~\cite{Alur1990MCRealTime,Alur1993DenseRealTime}.
An alternative approach to checking branching-time logic properties is the bottom-up algorithm~\cite{Alur1990MCRealTime,Clarke2018Handbook,Asarin1998SynthesisTimedAutomata}. This method begins by identifying all states that satisfy the relevant atomic propositions. It then iteratively propagates these results to compute states satisfying larger sub-properties until the original property is fully evaluated. While efficient when processing all reachable states is unavoidable, this approach may also waste time on unreachable states or irrelevant state-property combinations. Our on-the-fly algorithm mitigates these issues by computing only the necessary states and sub-properties on demand. Our new expansion abstraction relaxes the relevance criteria, bringing it closer to the bottom-up algorithm in this regard.

\paragraph*{Paper structure}
In \cref{sec:preliminaries}, we present the TMG formalism and TATL logic followed by the general EADG framework. Then we present our encoding of the TATL problem in EADGs in \cref{sec:encoding}. In \cref{sec:vertex_merging}, we introduce vertex merging to EADGs and an updated algorithm, allowing us to perform inclusion checking in our encoding, and we introduce our expansion abstraction. Finally, in \cref{sec:experiments}, we evaluate our algorithm and compare it against \uppaaltiga on the problem instances where it is possible.

\section{Preliminaries}\label{sec:preliminaries}

We shall first introduce the model of timed games with multiple players as well as a timed logics for describing properties of such systems. Afterwards, we present the extended abstract dependency graph framework.

\subsection{Timed Multiplayer Games}

Let $X$ be a finite set of real-valued variables called \emph{clocks}. Let $C(X)$ be the set of \emph{clock constraints} over the clocks $X$ generated by the following abstract syntax:
\[
    g::=x\bowtie k\mid x-y\bowtie k\mid g_1\land g_2
\]
where $k\in\Z$ and $x,y\in X$ and ${}\bowtie{}\in\{<,{}\leq{},=,{}\geq{},>\}$. Let $B(X)\subseteq C(X)$ be the subset that does not use any diagonal constraints of the form $x-y\bowtie k$. Let $\hat B(X)\subseteq B(X)$ be the subset that also does not use any constraints of the form $x<k$. A clock \emph{valuation} $\nu:X\to\R_{\geq 0}$ assigns each clock a real value. The valuation denoted $\veco$ assigns 0 to each clock. We write $\nu[Y]$ for a valuation that assigns 0 to any clock $x\in Y$ and assigns $\nu(x)$ to any clock $x\in X\setminus Y$. If $\delta\in\R_{\geq 0}$ then $\nu+\delta$ denotes the valuation such that $(\nu+\delta)(x)=\nu(v)+\delta$ for all $x\in X$. If $g\in C(X)$, we write $\nu\vDash g$ and say that $\nu$ satisfies $g$ if replacing all clocks in $g$ with their respective value in $\nu$ makes the expression evaluate to true.

\begin{definition}[Timed Automaton]
    A \emph{timed automaton (TA)} is a 6-tuple $\langle L, \ell_\mathrm{init}, X, A, T, I\rangle$ where $L$ is a finite set of locations, $\ell_\mathrm{init}\in L$ is the initial location, $A$ is a set of actions, $X$ is a finite set of real-valued clocks, $T\subseteq L\times B(X)\times A\times 2^X\times L$ is a finite set of edges each with a unique action from $A$, and $I:L\to\hat B(X)$ assigns an invariant to each location.
\end{definition}

The semantics of a TA $\mathcal A=\langle L, \ell_0, X, A, T, I\rangle$ can be described with a labeled transition system. The states are pairs $\langle \ell,\nu\rangle$ where $\ell\in L$ is a location and $\nu : X\to R_{\geq 0}$ is a valuation such that $\nu\vDash I(\ell)$ and $Q$ is the set of all states. Transition labels are discrete actions or real-valued delays, i.e.\ $A\cup \R_{\geq 0}$, and the transition relation $\to$ is a union of the binary relations $\xto a$ and $\xto\delta$ defined over states as follows:
\begin{itemize}
    \item $\langle\ell,\nu\rangle\xto a\langle\ell',\nu'\rangle$ if there exists an edge $\langle\ell,g,a,Y,\ell'\rangle\in T$ such that $v\vDash I(\ell)\land g$, $v'=v[Y]$, and $v'\vDash I(\ell')$, and
    \item $\langle\ell,\nu\rangle\xto\delta\langle\ell,\nu'\rangle$ if $\delta>0$, $v\vDash I(\ell)$, $v'=v+\delta$, and $v'\vDash I(\ell)$.
\end{itemize}

A \emph{run} in a TA is a sequence $\langle\ell_0,\nu_0\rangle\xto{t_0}\langle\ell_1,\nu_1\rangle\xto{t_1}\langle\ell_2,\nu_2\rangle\cdots$ such that $t_i\in A\cup\R_{\geq 0}$. A run is \textit{maximal} if it contains an infinite number of discrete action transitions, ends with an infinite number of delay transitions that sum to $\infty$ (divergence), or no transition is possible in the final state (a deadlock). For the latter case, there is always a well-defined final state since invariants of the form $x<k$ are disallowed. The set $Runs_\mathcal A(\langle \ell,\nu\rangle)$ contains all maximal runs starting from $\langle \ell,\nu\rangle$. Given a run $\sigma$ we write $\sigma[i]$ for the $i$th state $\langle\ell_i,\nu_i\rangle$.

\begin{definition}[Timed Multiplayer Game]
    A \emph{timed multiplayer game (TMG)} with $N$ players is a timed automaton where the actions $A$ have been partitioned into $N$ disjoint sets, $A=A_1\uplus\cdots\uplus A_N$. We denote by $\Sigma=\{1,\dotsc,N\}$ the set of players.
\end{definition}

\paragraph*{Strategies and outcomes}
A (memoryless) \emph{strategy} for a player $p\in\Sigma$ is a function $f_p:Q\to A_p\cup\{\lambda\}$ that maps each state to $\lambda$ or a discrete action that is enabled in the state and belongs to $p$. The strategy informs $p$ what to do in the given state and $\lambda$ is a special symbol indicating that player $p$ should do nothing and wait. For states where a delay transition is not possible, a strategy can only output $\lambda$ if the player has no enabled actions available. Let $F_p$ be the set of all strategies for the player $p$. By $\xi_S$ we denote a strategy profile that assigns to each player $p\in S\subseteq\Sigma$ a strategy $f_p\in F_p$, and $\Xi(S)$ is the set of all such strategy profiles over $S$. When players pick actions according to their strategies, it restricts the possible runs. We write $Out_\mathcal A(\xi_S,\langle\ell,\nu\rangle)\subseteq Runs_\mathcal A(\langle\ell,\nu\rangle)$ for the subset of \emph{outcomes} (runs) induced when the players of $S$ adhere to the strategies of $\xi_S$. 

\begin{definition}[Outcomes]
    Let $\mathcal A=\langle L, \ell_\mathrm{init}, X, A, T, I\rangle$ be a TMG with players $\Sigma=\{1,\dotsc,N\}$, let $\langle\ell_0,\nu_0\rangle$ be a state where $\nu_0\vDash I(\ell_0)$, and let $\xi_S\in\Xi(S)$ be a strategy profile for $S\subseteq\Sigma$. A maximal run $\sigma=\langle\ell_0,\nu_0\rangle\xto{t_0}\langle\ell_1,\nu_1\rangle\xto{t_1}\langle\ell_2,\nu_2\rangle\cdots\in Runs_\mathcal A(\langle\ell_0,\nu_0\rangle)$ is in $Out_\mathcal A(\xi_S,\langle\ell_0,\nu_0\rangle)$ iff the following conditions hold:
    \begin{itemize}
        \item for all $i\geq 0$ such that $t_i=\delta\in\R_{\geq 0}$ we have that $\xi_S(p)(\langle\ell_i,\nu_i+\delta'\rangle)=\lambda$ for all $0\leq\delta'<\delta$ and all $p\in S$,
        \item for all $i\geq 0$ such that $t_i=a\in A$ we have either $a\notin \bigcup_{p\in S} A_p$ or $\xi_S(p)(\langle\ell_i,\nu_i\rangle)=a$ with $a\in A_p$ for some $p\in S$.
    \end{itemize}
\end{definition}

Note that we allow Zeno behavior (an infinite subsequence consisting of no delays) from both the coalition $S$ and the opposition $\overline S$.

\subsection{Timed Alternating-Time Temporal Logic}

The alternating-time temporal logic~\cite{Rajeev2002AtlLogics} is a logic that offers quantification over possible outcomes resulting from a coalition of players in a multiplayer game.

\begin{definition}[Alternating-Time Temporal Logic]
    Given a set of atomic propositions $\Pi$, a~set of clocks $X$, and a set of players $\Sigma$, an \emph{alternating-time temporal logic (ATL)} property conforms to the abstract syntax:
    \[
        \phi ::= \pi\mid g\mid\neg\phi_1\mid\phi_1\lor\phi_2\mid\coal S\bigcirc\phi_1\mid\coal S(\phi_1\until\phi_2)\mid\coalbr S(\phi_1\until\phi_2)
    \]
    where $\pi\in\Pi$, $g\in C(X)$, and $S\subseteq\Sigma$ is a coalition of players.
\end{definition}

Given a TMG $\mathcal A$ with players $\Sigma=\{1,\dotsc,N\}$ and a function $Lab:L\to 2^\Pi$ labeling locations with atomic propositions, the satisfaction relation $\vDash$ over states and ATL properties is defined inductively as follows:
\begin{itemize}
    \item $\langle\ell,\nu\rangle\vDash\pi$ iff $\pi\in Lab(\ell)$,
    \item $\langle\ell,\nu\rangle\vDash g$ iff $\nu\vDash g$,
    \item $\langle\ell,\nu\rangle\vDash\neg\phi$ iff $\langle\ell,\nu\rangle\not\vDash\phi$,
    \item $\langle\ell,\nu\rangle\vDash\phi_1\lor\phi_2$ iff $\langle\ell,\nu\rangle\vDash\phi_1$ or $\langle\ell,\nu\rangle\vDash\phi_2$,
    \item $\langle\ell,\nu\rangle\vDash\coal S\bigcirc\phi_1$ iff there exists a strategy profile $\xi_S\in\Xi(S)$ such that for all maximal runs $\sigma=\langle\ell_0,\nu_0\rangle\xto{t_0}\langle\ell_1,\nu_1\rangle\xto{t_1}\langle\ell_2,\nu_2\rangle\cdots\in Out_\mathcal A(\xi_S,\langle\ell,\nu\rangle)$ we have $\sigma[i+1]\vDash\phi_1$ where $i\geq 0$ is the smallest $i$ such that $t_i\in A$ and such an $i$ must exist,\footnote{The next operator is usually left out of timed systems, but it is well-defined with action-based semantics like here.}
    \item $\langle\ell,\nu\rangle\vDash\coal S(\phi_1\until\phi_2)$ iff there exists a strategy profile $\xi_S\in\Xi(S)$ such that for all maximal runs $\sigma\in Out_\mathcal A(\xi_S,\langle\ell,\nu\rangle)$ we have $\sigma\vDash_r\phi_1\until\phi_2$ (defined below),
    \item $\langle\ell,\nu\rangle\vDash\coalbr S(\phi_1\until\phi_2)$ iff for all strategy profiles $\xi_S\in\Xi(S)$ there exists a maximal run $\sigma\in Out_\mathcal A(\xi_S,\langle\ell,\nu\rangle)$ such that $\sigma\vDash_r\phi_1\until\phi_2$,
\end{itemize}

\noindent and for a maximal run $\sigma=\langle\ell_0,\nu_0\rangle\xto{t_0}\langle\ell_1,\nu_1\rangle\xto{t_1}\langle\ell_2,\nu_2\rangle\cdots$ we have $\sigma\vDash_r\phi_1\until\phi_2$ iff there exists an $i\geq 0$ such that:
\begin{itemize}
    \item for all $j< i$:
    \begin{itemize}
        \item if $t_j\in\R_{\geq 0}$ then for all $\delta\in [0,t_j)$ we have $\langle\ell_j,\nu_j+\delta\rangle\vDash\phi_1$, or
        \item if $t_j\in A$ instead then $\langle\ell_j,\nu_j\rangle\vDash\phi_1$, and
    \end{itemize}
    \item either $\langle\ell_i,\nu_i\rangle\vDash\phi_2$ or $t_{i}\in\R_{\geq 0}$ and there exists a $\delta\in[0,t_i)$ such that $\langle\ell_i,\nu_i+\delta\rangle\vDash\phi_2$, and for all $\delta'\in[0,\delta)$ we have $\langle\ell_i,\nu_i+\delta'\rangle\vDash\phi_1\lor\phi_2$. We note that the disjunction $\phi_1\lor\phi_2$ is necessary for timed big-step runs (instead of simply $\phi_1$). See \cite{Anderson2007VerificationRTS} for details.
\end{itemize}

We note that $\coalbr{\emptyset}\equiv\exists$ and $\coal{\emptyset}\equiv\forall$ and thus ATL is a superset of CTL~\cite{Rajeev2002AtlLogics}. Other notable abbreviations and equivalences are $\pi\lor\neg\pi\equiv\mathbf{true}$ and $\coal S(\mathbf{true}\,\until\phi)\equiv\coal S\Diamond\phi$ and $\neg\coal S\Diamond\phi\equiv\coalbr S\Box\neg\phi$ and $\neg\coal S\bigcirc\phi\equiv\coalbr S\bigcirc\neg\phi$.

\begin{definition}[Timed ATL]
    \emph{Timed alternating-time temporal logic (TATL)} is the timed extension of ATL and is defined as ATL but with a new freeze operator: $z.\phi$, where $z\in X$ is a \emph{formula clock} and $\phi$ is another TATL property. Note that a formula clock $z$ may not be used in the guards and invariants of the TMG. Given a state $\langle\ell,\nu\rangle\in Q$ we have $\langle\ell,\nu\rangle\vDash z.\phi$ iff $\langle\ell,\nu[z]\rangle\vDash\phi$.
\end{definition}

The TATL freeze operator is commonly used indirectly through the more intuitive temporal operator $\until_{\lhd k}$ (where $\lhd\in\{<,\leq\})$. For example, if $z$ is a new clock that does not appear elsewhere in the model or the formula then:
\[
    \coal S(\phi_1\until_{\lhd k}\phi_2)\equiv z.\coal S((\phi_1\land z\lhd k)\,\until\phi_2)
\]
meaning that coalition $S\subseteq\Sigma$ has a strategy profile such that when adhered to, the system always reaches a state satisfying $\phi_2$ (strictly) within $k$ time units by a path where $\phi_1$ invariantly holds.

An example strategy profile that witnesses $\mathcal A\vDash\coal{\ExamPlayerI,\ExamPlayerIII}(\neg\mathbf{C}\;\until\;\mathbf{Goal})$ for the TMG $\mathcal A$ in \cref{fig:intro_example} is given in \cref{fig:intro_example:strategy_profile}.
Since some states are never visited under this strategy, a partial strategy profile would also suffice as a witness. Such partial strategy witnesses enable our algorithm to terminate early for positive cases.

\subsection{Extended Abstract Dependency Graphs}\label{sec:prelim_eadg}

A \textit{partially ordered set} $\langle\mathcal D,\sqsubseteq\rangle$ is
a set $\mathcal D$ together with a binary relation $\sqsubseteq\,\subseteq \mathcal D \times \mathcal D$ that is reflexive, transitive and anti-symmetric.
Given partially ordered sets $\langle\mathcal D_1, \sqsubseteq_1\rangle$ and $\langle\mathcal D_2, \sqsubseteq_2\rangle$, a function $f:\mathcal D_1\to\mathcal D_2$ is \textit{monotonically increasing} if $d\sqsubseteq_1 d'$ implies $f(d)\sqsubseteq_2 f(d')$ and \textit{monotonically decreasing} if $d\sqsubseteq_1 d'$ implies $f(d)\sqsupseteq_2 f(d')$. When we just say \textit{monotonic} we refer to increasing monotonicity.
When a function has multiple inputs, monotonicity is with respect to a specific input. Unless we specify further, our monotonic multi-input functions are monotonic with respect to all inputs.

\begin{definition}[Noetherian Ordering Relation]\label{def:nor}
    A \emph{Noetherian ordering relation with a least element (NOR)} is a triple $\langle\mathcal D, \sqsubseteq, \bot\rangle$ such that $\langle\mathcal D,\sqsubseteq\rangle$ is a partially ordered set, $\bot\in\mathcal D$ is the least element such that $\bot\sqsubseteq d$ for all $d\in\mathcal D$, and $\sqsubseteq$ satisfies the stabilizing ascending chain condition: for any chain $d_1\sqsubseteq d_2\sqsubseteq d_3\sqsubseteq\cdots$ there exists a point $i$ such that $d_i=d_j$ for all $j>i$.
\end{definition}

\begin{definition}[Abstract Dependency Graph]\label{def:adg}
    An \emph{abstract dependency graph (ADG)} is tuple $\langle V, E, \mathcal D, \mathcal E\rangle$ where $V$ is a finite set of vertices, $E:V\to V^*$ is an edge function from vertices to strings of vertices, $\mathcal D$ is a NOR, and $\mathcal E(v):\mathcal D^n\to\mathcal D$ is a monotonic value function at vertex $v\in V$ that takes $n$ arguments where $n=|E(v)|$.
\end{definition}

We write $v\to v'$ if $v,v'\in V$ and $v'\in E(v)$, and $\to^+$ denotes the transitive closure of $\to$. The empty string of vertices is denoted $\varepsilon$. Vertices $v$ where $E(v)=\varepsilon$ have constant value functions by definition.

\begin{definition}[Extended Abstract Dependency Graph]\label{def:eadg}
    An \emph{extended abstract dependency graph (EADG)} is a tuple $\langle V, E, \mathcal D, \mathcal E\rangle$ where $V$, $E$, and $\mathcal D$ are defined as for ADGs, but now $\mathcal E(v):\mathcal D^n\to\mathcal D$ is a (possibly non-monotonic) value function at vertex $v\in V$ that takes $n$ arguments with $n=|E(v)|$. Furthermore, if $\mathcal E(v)$ is non-monotonic, then $v$ is not in a cycle, i.e.\ it is not the case that $v\to^+ v$.
\end{definition}

\begin{figure}
    \centering
    \begin{subfigure}{0.36\textwidth}
        \centering
        \begin{tikzpicture}[
                node distance=0.8,
                input/.style = {draw,circle,minimum size=1mm,inner sep=0pt, outer sep=0pt},
                efunc/.style = {inner sep=0.25pt},
                dep/.style = {-stealth}
            ]
            \node[draw] (A) {$A$};
            \node[efunc, below=0 of A] (Ae) {\tiny$\max(\;\;,\;\;)$};
            \node[input] (A1) at ($(A.south) + (1mm,-1mm)$) {};
            \node[input] (A2) at ($(A.south) + (3.33mm,-1mm)$) {};
            
            \node[draw, below left=0.5 and 0.3 of A] (B) {$B$};
            \node[efunc, below=0 of B] (Be) {\tiny$-$};
            \node[input] (B1) at ($(B.south) + (-2mm,-1mm)$) {};
            \node[input] (B2) at ($(B.south) + (2mm,-1mm)$) {};
            
            \node[draw, below right=0.5 and 0.3 of A] (C) {$C$};
            \node[efunc, below=0 of C] (Ce) {\tiny$\min(\;\;,\;\;)$};
            \node[input] (C1) at ($(C.south) + (0.66mm,-1mm)$) {};
            \node[input] (C2) at ($(C.south) + (3mm,-1mm)$) {};
            
            \node[draw, below left=0.5 and 0.3 of B] (D) {$D$};
            \node[efunc, below=0 of D] (De) {\tiny$4+\;\;$};
            \node[input] (D1) at ($(D.south) + (2mm,-1mm)$) {};
            
            \node[draw, below right=0.5 and 0.3 of B] (E) {$E$};
            \node[efunc, below=0 of E] (Ee) {\tiny$\max(\;\;,\;\;+3)$};
            \node[input] (E1) at ($(E.south) + (-0.7mm,-1mm)$) {};
            \node[input] (E2) at ($(E.south) + (2.1mm,-1mm)$) {};
            
            \node[draw, below right=1 and 0 of C] (F) {$F$};
            \node[efunc, below=0 of F] (Fe) {\tiny$3$};
            
            \node[draw, below=of D] (G) {$G$};
            \node[efunc, below=0 of G] (Ge) {\tiny$10$};
            
            \node[draw, below=of E] (H) {$H$};
            \node[efunc, below=0 of H] (He) {\tiny$2$};

            \node[fit=(D)(E)(F)(G)(H),draw,dashed,gray,inner sep=2mm] (Comp0box) {};
            \node[above left=0 of Comp0box.south east] (Comp0) {$C_0$}; 

            \node[fit=(Comp0box)(A)(B)(C),draw,dashed,gray,inner sep=2mm] (Comp1box) {};
            \node[below left=0 of Comp1box.north east] (Comp1) {$C_1$};
    
            \draw[dep] (A1) -- (B);
            \draw[dep] (A2) -- (C);
            \draw[dep] (B1) -- (D);
            \draw[dep] (B2) -- (E);
            \draw[dep] (C1) -- (F);
            \draw[dep] (C2) edge [out=45, in=0, looseness=1.5] (A); %
            \draw[dep] (D1) -- (G);
            \draw[dep] (E1) -- (H);
            \draw[dep] (E2) edge [out=300, in=190, looseness=1.5] (F);
        \end{tikzpicture}
        \caption{Example EADG}
        \label{fig:eadge_example:pre_merge}
    \end{subfigure}\;%
    \begin{subfigure}{0.26\textwidth}
        \centering
        \begin{tabular}{c|cc}
            $v$ & $\alpha_{\min}^{C_0}$ & $\alpha_{\min}^{C_1}$ \\ \hline
            A & - & 8 \\
            B & - & 8 \\
            C & - & 3 \\
            D & 14 & 14 \\
            E & 6 & 6 \\
            F & 3 & 3 \\
            G & 10 & 10 \\
            H & 2 & 2 \\
        \end{tabular}
        \caption{Fixed points of components $C_0$ and $C_1$}
        \label{fig:eadge_example:alpha}
    \end{subfigure}\;\;%
    \begin{subfigure}{0.36\textwidth}
        \centering
        \begin{tikzpicture}[
                node distance=0.8,
                input/.style = {draw,circle,minimum size=1mm,inner sep=0pt, outer sep=0pt},
                efunc/.style = {inner sep=0.25pt},
                dep/.style = {-stealth}
            ]
            \node[draw] (A) {$A$};
            \node[efunc, below=0 of A] (Ae) {\tiny$\max(\;\;,\;\;)$};
            \node[input] (A1) at ($(A.south) + (1mm,-1mm)$) {};
            \node[input] (A2) at ($(A.south) + (3.33mm,-1mm)$) {};
            
            \node[draw, below left=0.5 and 0.3 of A] (B) {$B$};
            \node[efunc, below=0 of B] (Be) {\tiny$-f(\;\;)$};
            \node[input] (B1) at ($(B.south) + (-4mm,-1mm)$) {};
            \node[input] (B2) at ($(B.south) + (1.7mm,-1mm)$) {};
            
            \node[draw, below right=0.5 and 0.3 of A] (C) {$C$};
            \node[efunc, below=0 of C] (Ce) {\tiny$\min(\;\;,\;\;)$};
            \node[input] (C1) at ($(C.south) + (0.66mm,-1mm)$) {};
            \node[input] (C2) at ($(C.south) + (3mm,-1mm)$) {};
            
            \node[draw, below left=0.5 and 0.3 of B] (D) {$D$};
            \node[efunc, below=0 of D] (De) {\tiny$4+\;\;$};
            \node[input] (D1) at ($(D.south) + (2mm,-1mm)$) {};
            
            \node[draw, below right=1 and 0 of C] (F) {$F$};
            \node[efunc, below=0 of F] (Fe) {\tiny$3$};

            \node[text width=30mm, below right=0.35 and -0.6 of B] {\tiny\shortstack{where\\$f(x)=x+3$}};
            
            \node[draw, below=0.5 and 0.3 of D] (G) {$G$};
            \node[efunc, below=0 of G] (Ge) {\tiny$10$};
            
            \node[draw, below=0.5 and 0.3 of E] (H) {$H$};
            \node[efunc, below=0 of H] (He) {\tiny$2$};
            \draw (H.north east) -- (H.south west);
            \draw (H.north west) -- (H.south east);
    
            \draw[dep] (A1) -- (B);
            \draw[dep] (A2) -- (C);
            \draw[dep] (B1) -- (D);
            \draw[dep] (B2) -- (F);
            \draw[dep] (C1) -- (F);
            \draw[dep] (C2) edge [out=45, in=0, looseness=1.5] (A); %
            \draw[dep] (D1) -- (G);
        \end{tikzpicture}
        \caption{The EADG after merge by $E\preceq_f F$ where $f(x)=x+3$}
        \label{fig:eadge_example:post_merge}
    \end{subfigure}%
    \caption{ (a) An example EADG $\mathcal G$ over $\mathcal D=\langle\N_0\cup\{\infty\},\geq,\infty\rangle$. 
    For each vertex $v$, the value function $\mathcal E(v)$ is displayed below the vertex, and the edges $E(v)$ are displayed using small circles inside the value functions, indicating the arity of the value function as well as which vertex assignment is used as input. Leaf nodes $F$, $G$, and $H$ have no dependencies and their value function is a constant function. Since $\mathcal E(B)$ is non-monotonic, the graph has two components $C_0$ and $C_1$ which are highlighted with dashed boxes. (b) The fixed-point assignments of component $C_0$ and $C_1$ found by a fixed-point computation. (c) The EADG $\mathcal G[E\mapsto_f F]$ where vertex $E$ has been removed through derivation $E\preceq_f F$. The value function of vertex $B$ is pointwise composed with the derive function $f$ on the second input which now points to the assignment of $F$ instead. Vertex $H$ now has no dependents and can be pruned.}
    \label{fig:eadg_example}
\end{figure}
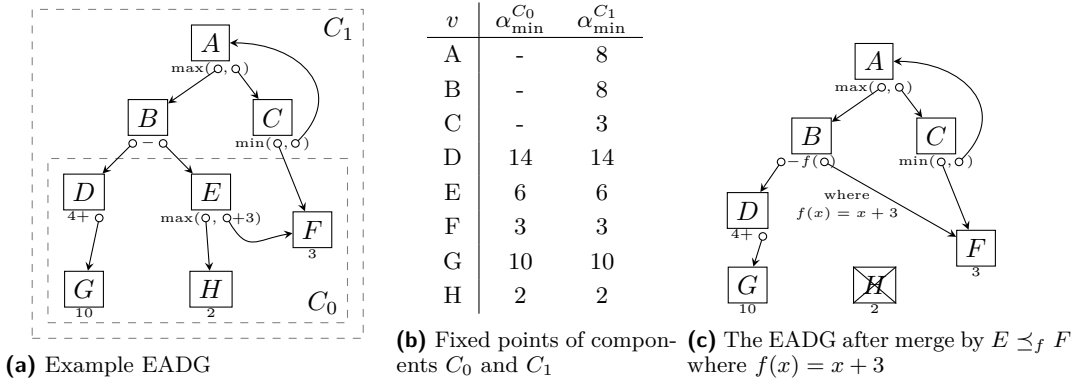

EADGs can be used to encode and solve various problems. Each vertex in the graph represents a problem and the NOR values $\mathcal D$ represent possible answers to problems with increasing accuracy. The value function at each vertex describes how the answer to the problem represented by the vertex can be derived from the answers to other (sub)problems. To describe this formally, we must partition the graph into components. Consider the function:
\begin{align}
    dist(v)=\max\{m\in\N\mid\ & \exists v_0v_1\cdots v_n\in V^*, v=v_0, \forall i\in\{1,\dotsc,n\}, v_{i-1}\to v_i,\\&m=|\{v_i\mid \mathcal E(v_i)\textrm{ is non-monotonic}\}|\} \nonumber
\end{align}
describing how many non-monotonic value functions a vertex depends on. Since an EADG has no cycles with non-monotonic functions, $dist(v)$ is well-defined. The subgraphs induced by $dist$ are called components and $C_i\subseteq V$ is the $i$th component where $dist(v)\leq i$ for all $v\in C_i$. Component $C_0$ depends only on monotonic value functions.

Let $\alpha:V\to\mathcal D$ be an assignment from vertices to NOR values with $\alpha_\bot$ being the assignment such that $\alpha_\bot(v)=\bot$ for all vertices $v\in V$.
Let $F_0$ be an update function such that:
\begin{align}
    F_0(\alpha)(v)=\mathcal E(v)(\alpha(v_1),\alpha(v_2),\dotsc,\alpha(v_n))
\end{align}
where $v\in C_0$ and $E(v)=v_1 v_2\dots v_n$. Since all value functions in $C_0$ are monotonic and $\mathcal D$ is a NOR, repeated application of $F_0$ on $\alpha_\bot$ eventually reaches a least fixed point, i.e. there exists an $m>0$ such that $F_0^m(\alpha_\bot)=F_0^{m+1}(\alpha_\bot)$. We denote this fixed point $\alpha_{\min}^{C_0}$. For each component $C_i$ where $i>0$ the update function $F_i$ is defined as follows:
\begin{align}
    F_i(\alpha)(v)=\begin{cases}
        \mathcal E(v)(\alpha(v_1),\alpha(v_2),\dotsc,\alpha(v_n)) & \textrm{ if }\mathcal E(v)\textrm{ is monotonic}\\
        \mathcal E(v)(\alpha_{\min}^{C_{i-1}}(v_1),\alpha_{\min}^{C_{i-1}}(v_2),\dotsc,\alpha_{\min}^{C_{i-1}}(v_n)) & \textrm{ otherwise}
    \end{cases}
\end{align}
where $E(v)=v_1 v_2\dots v_n$ and the assignment $\alpha_{\min}^{C_{i-1}}$ is the minimal fixed point on component $C_{i-1}$, i.e.\ the fixed points are defined inductively \cite{Enevoldsen2022EADGs}. Finally, let $C_{\max}$ be the component associated with the greatest $dist$ in EADG $G$, and let $\alpha_{\min}^G$ denote $\alpha_{\min}^{C_{\max}}$.
Given a vertex $v_0\in V$, the value $\alpha_{\min}^G(v_0)$ can be efficiently found using the sound and complete on-the-fly algorithm introduced in \cite{Enevoldsen2022EADGs}.

A small example EADG can be seen in \cref{fig:eadge_example:pre_merge} with the NOR $\mathcal D=\langle\N_0\cup\{\infty\},\geq,\infty\rangle$. In this domain, chains are descending, stabilizing before or at 0. In \cref{fig:eadge_example:alpha} we show the minimum fixed-point assignment of the components in the example EADG.

\section{Encoding TATL to EADGs}\label{sec:encoding}

In this section we will introduce our symbolic encoding of the TATL problem to EADGs.

\paragraph*{Symbolic Operations on Zones and Federations}
Due to the uncountable size of the state space, automated verification of timed automata groups the concrete states into effectively representable sets of states and valuations~\cite{Alur1990MCRealTime,Bengtsson2004TASemanticsAlgoTools}. We shall now define some useful operators on such sets. Let $\coalbr g=\{\nu\in\R_{\geq 0}^X\mid \nu\vDash g\}$ be the set of all valuations satisfying $g\in C(X)$. A \emph{zone} is a set $Z\subseteq\R_{\geq 0}^X$ of valuations where for some $g\in C(X)$ we have $\coalbr g=Z$. A finite union of zones is called a federation. If $W\subseteq\R_{\geq 0}^X$ is a set of valuations, then:
\begin{itemize}
    \item $W^\nearrow=\{\nu+\delta\mid\nu\in W,\delta\in\R_{\geq 0}\}$ are the timed successors of $W$,
    \item $W^\swarrow=\{\nu'\in\R_{\geq 0}^X\mid\exists\nu\in W,\exists\delta\in\R_{\geq 0},\nu'+\delta=\nu\}$ are the timed predecessors of $W$,
    \item $W[Y]=\{\nu[Y]\mid\nu\in W\}$ is a reset of the clocks $Y\subseteq X$ in $W$,
    \item $W\#x=\{\nu'\in\R_{\geq 0}^X\mid\exists\nu\in W, \forall y\in X, y\neq x\implies \nu(y)=\nu'(y)\}$  frees clock $x\in X$ in $W$. 
\end{itemize}
Zones are closed under all the above operations~\cite{Bengtsson2004TASemanticsAlgoTools}. Now consider a set of states $Q' \subseteq Q$ of a timed automaton $\langle L,\ell_\text{init},X,A,T,I\rangle$. We extend these operations on clock valuations to the sets of states such that if $F\in\{\cdot^\nearrow,\cdot^\swarrow,\cdot[Y],\cdot\#x\}$ then $F(Q')=\{\langle\ell,\nu'\rangle\mid\exists\langle\ell,\nu\rangle\in Q', \nu'\in F(\{\nu\})\cap \coalbr{I(\ell)}\}$. Given an action $a\in A$, we define the discrete $a$-predecessors and $a$-successors as $Pred_a(Q')=\{\langle\ell,\nu\rangle\mid\exists\langle\ell',\nu'\rangle\in Q',\langle\ell,\nu\rangle\xto a\langle\ell',\nu'\rangle\}$ and $Post_a(Q')=\{\langle\ell,\nu\rangle\mid\exists\langle\ell',\nu'\rangle\in Q',\langle\ell',\nu'\rangle\xto a\langle\ell,\nu\rangle\}$, respectively. Both $Pred_a$ and $Post_a$ preserve zones and federations~\cite{Bengtsson2004TASemanticsAlgoTools}.
When the automata is a TMG with players $\Sigma=\{1,\dotsc,N\}$ and $S\subseteq\Sigma$ is a coalition, then we let $A_S$ denote $\bigcup_{p\in S} A_p$. The $S$-controllable discrete predecessors of the coalition is defined by $Pred_S(Q')=\bigcup_{a\in A_S} Pred_a(Q')$. Similarly, the $S$-uncontrolled discrete predecessors is $Pred_{\overline S} (Q')$ where ${\overline S}=\Sigma\setminus S$. Safe timed predecessors are those that avoid a set of states $Q''\subseteq Q$ even as time elapses. As defined in~\cite{Cassez2005TimedGames}:
\[
    Pred_\lambda(Q', Q'')=\{\langle\ell,\nu-\delta\rangle\mid\langle\ell,\nu\rangle\in Q',\delta\geq 0,\forall\delta'\in[0,\delta],\langle\ell,\nu-\delta'\rangle\notin Q''\} \ .
\]
The states of $Q'$ where time cannot pass are defined as $Q'^\nodelay=\{\langle\ell,\nu\rangle\in Q'\mid\exists"x\leq k"\in I(\ell),\nu(x)=k\}$. These states are also called \emph{time-locked}. We note that the $\cdot^{\nodelay}$ operator preserves federations. Finally, let $Zone(Q)\subseteq 2^Q$ be all subsets of $Q$ that can be represented with a location-zone pair $\langle\ell,Z\rangle$ (also called a~symbolic state), and let $Fed(Q)\subseteq 2^Q$ be all subsets of $Q$ that can be represented with a location-federation pair $\langle\ell,\bigcup_i Z_i\rangle$.

\paragraph*{The Encoding}

We shall now encode the TATL problem on TMGs as an EADG $G$ in order to answer the question: given a TMG $\mathcal A=\langle L, \ell_\mathrm{init}, X, A, T, I\rangle$ with players $\Sigma=\{1,\dotsc,N\}$ and a TATL formula $\phi_0$, is it the case that $\langle\ell_\textrm{init},\veco\rangle\vDash\phi_0$?
The vertices in EADG will have the form $\langle R,\phi\rangle$ where $R\in Zone(Q)$ is a symbolic state and $\phi$ is a TATL formula (a sub-formula of $\phi_0$). We shall sometimes specify the location and zone of $R$ and write $\langle\ell,Z,\phi\rangle$ for vertices instead, but the notation is often simpler for sets of states so we use $R$ in those cases. Our root vertex is $\langle\ell_\textrm{init},\{\veco\}^\nearrow\cap\coalbr{I(\ell_\textrm{init})}, \phi_0\rangle$ and all dependencies of all vertices are generated using operators that preserve symbolic states. Without loss of generality, we assume all clocks are bounded\footnote{It is also possible to use extrapolation techniques instead \cite{Behrmann2003Extrap}.}~\cite{Cassez2005TimedGames,Bengtsson2004TASemanticsAlgoTools} and hence there are only finitely many symbolic states in practice, and therefore the dependency graph is finite.
Our assignment domain is the NOR $\langle Fed(Q),\subseteq,\emptyset\rangle$ and we typically denote elements from our NOR using $W$. We restrict ourselves such that $\alpha_{\min}^G(\langle R,\phi\rangle)\subseteq R$ by construction and our goal is that $\langle\ell,\nu\rangle\in\alpha_{\min}^G(\langle R,\phi\rangle)$ iff $\langle\ell,\nu\rangle\vDash\phi$. 
The value functions of our EADG rely on the helper function $\mathit{Forceable}_S(W_{\phi_1}, W_{\phi_2}, \mathcal W)$ and its counterpart $\mathit{Unavoidable}_S(W_{\phi_1}, W_{\phi_2}, \mathcal W)$ where $S$ is a coalition of players, $W_{\phi_1}$ and $W_{\phi_2}$ are sets of states (typically associated with two TATL formulae), and $\mathcal W$ is a set of states.
We have $\langle\ell,\nu\rangle\in \mathit{Forceable}_S(W_{\phi_1}, W_{\phi_2}, \mathcal W)$ if and only if there exists a $\delta\geq 0$ such that $\langle\ell,\nu\rangle\xto\delta\langle\ell,\nu+\delta\rangle$ and 
\begin{itemize}
    \item for all $\delta'\in[0,\delta]$ either $\langle\ell,\nu+\delta'\rangle\in W_{\phi_2}$, or $\langle\ell,\nu+\delta'\rangle\in W_{\phi_1}$ and for all $a_i\in A_{\overline S}$ if $\langle\ell,\nu+\delta'\rangle\xto{a_i}\langle\ell',\nu'\rangle$ then $\langle\ell',\nu'\rangle\in\mathcal W$,
    \item and additionally
    \begin{itemize}
        \item $\langle\ell,\nu+\delta\rangle\in W_{\phi_2}$, or
        \item $\langle\ell,\nu+\delta\rangle\xto{a_j} \langle\ell',\nu'\rangle\in\mathcal W$ for some $a_j\in A_S$, or
        \item $\langle\ell,\nu+\delta\rangle\not\xto{\delta''}$ for any $\delta''>0$ and $\langle\ell,\nu+\delta\rangle\xto{a_j}\langle\ell',\nu'\rangle\in\mathcal W$ for some $a_j\in A_{\overline S}$.
    \end{itemize}
\end{itemize}
That is, the set is the subset of $W_{\phi_1}\cup W_{\phi_2}$ where coalition $S$ can controllably get to $W_{\phi_2}$ with delays or to $\mathcal W$ with delays and a discrete action all while staying in $W_{\phi_1}$ until then. The set can be described symbolically as follows:
\begin{align}\label{eq:forceable}
    & \mathit{Forceable}_S(W_{\phi_1},W_{\phi_2}, \mathcal W) =\nonumber\\
    &\quad Pred_\lambda\left(W_{\phi_2}\cup Pred_S(\mathcal W)\cup\mathcal H, \left[\overline{W_{\phi_1}}\cup Pred_{\overline{S}}(\overline{\mathcal W})\right]\setminus W_{\phi_2}\right)\\
    \label{eq:forceable:H}
    &\text{where } \mathcal H =Q^{\nodelay}\cap Pred_{\overline S}(\mathcal W)\setminus Pred_\Sigma(\overline{\mathcal W}).
\end{align}%
Similarly, we have $\langle\ell,\nu\rangle\in \mathit{Unavoidable}(W_{\phi_1}, W_{\phi_2}, \mathcal W)$ if and only if
there exists a $\delta\geq 0$ such that $\langle\ell,\nu\rangle\xto\delta\langle\ell,\nu+\delta\rangle$ and
\begin{itemize}
    \item for all $\delta'\in[0,\delta)$ we have $\langle\ell,\nu+\delta'\rangle\in W_{\phi_1}$ and for all $a_i\in A_S$ if $\langle\ell,\nu+\delta'\rangle\xto{a_i}\langle\ell',\nu'\rangle$ then $\langle\ell',\nu'\rangle\in\mathcal W$,
    \item and additionally
    \begin{itemize}
        \item $\langle\ell,\nu+\delta\rangle\in W_{\phi_2}$, or
        \item $\langle\ell,\nu+\delta\rangle\in W_{\phi_1}$ and $\langle\ell,\nu+\delta\rangle\xto{a_j} \langle\ell',\nu'\rangle\in\mathcal W$ for some $a_j\in A_{\overline S}$, or
        \item $\langle\ell,\nu+\delta\rangle\in W_{\phi_1}$ and $\langle\ell,\nu+\delta\rangle\not\xto{\delta''}$ for any $\delta''>0$ and for all $a_i\in A_S$ if $\langle\ell,\nu+\delta'\rangle\xto{a_i}\langle\ell',\nu'\rangle$ then $\langle\ell',\nu'\rangle\in\mathcal W$ and there exists at least one $a_i\in A_S$ such that $\langle\ell,\nu+\delta\rangle\xto{a_i}$.
    \end{itemize}
\end{itemize}
That is, the set contains the subset of $W_{\phi_1}\cup W_{\phi_2}$ where coalition $S$ cannot avoid that the system stays in $W_{\phi_1}$ and eventually reaches $W_{\phi_2}$ through delays, or reaches $\mathcal W$ through a~delay followed by a discrete action.
The set can be described symbolically as follows:
\begin{align}\label{eq:avoidable}
    & \mathit{Unavoidable}_S(W_{\phi_1}, W_{\phi_2}, \mathcal W)=\nonumber\\
    &\quad Pred_\lambda\left(W_{\phi_2}\cup Pred_{\overline{S}}(\mathcal W)\cup\mathcal H, \left[\overline{W_{\phi_1}}\cup (Pred_S(\overline{\mathcal W})\setminus Pred_{\overline{S}}(\mathcal W))\right]\setminus W_{\phi_2}\right)\\
    \label{eq:unavoidable:H}
    &\text{where } \mathcal H=Q^{\nodelay}\cap Pred_S(\mathcal W).
\end{align}%
Assuming a fixed ordering of actions $A=\{a_1,\dots,a_n\}$, we now define $E(v)$ and $\mathcal E(v)$ of the EADG as follows based on the form of vertex $v$:

\begin{itemize}
    \item Case $v=\langle R,\pi\rangle$:\quad $E(v)=\varepsilon$ \quad and\quad $\mathcal E(v)(\varepsilon)=\{\langle\ell,\nu\rangle\in R\mid \pi\in Lab(\ell)\}$.

    \item Case $v=\langle R,g\rangle$:\quad $E(v)=\varepsilon$ \quad and\quad $\mathcal E(v)(\varepsilon)=\{\langle\ell,\nu\rangle\in R\mid\nu\in\coalbr{g}\}$.

    \item Case $v=\langle R,\neg\phi\rangle$:\quad $E(v)=\langle R,\phi\rangle$ \quad and\quad $\mathcal E(v)(W)=R\setminus W$.

    \item Case $v=\langle R,\phi_1\lor\phi_2\rangle$: $E(v)=\langle R,\phi_1\rangle\langle R,\phi_2\rangle$
            and $\mathcal E(v)(W_1,W_2)=W_1\cup W_2$.

    \item Case $v=\langle R,\phi_1\land\phi_2\rangle$: $E(v)=\langle R,\phi_1\rangle\langle R,\phi_2\rangle$
            and $\mathcal E(v)(W_1,W_2)=W_1\cap W_2$.

    \item Case $v=\langle R,z.\phi\rangle$: $E(v)=\langle R[z]^\nearrow,\phi\rangle$
    and $\mathcal E(v)(W)=(W\cap Q[z])\#z\cap R$.

    \item Case $v=\langle R,\coal S\bigcirc\phi\rangle$:
        \begin{align*}
            & E(v)=v_{a_1}\dots v_{a_n}\quad \textrm{where } v_{a_i} = \langle Post_{a_i}(R)^\nearrow,\phi\rangle \\
            & \mathcal E(v)(W_{a_1},\dotsc,W_{a_n}) = \mathit{Forceable}_S(R,\emptyset, \bigcup_i W_{a_i})
        \end{align*}

    \item Case $v=\langle R,\coal S(\phi_1\until\phi_2)\rangle$:
        \begin{align*}
            & E(v)=\langle R,\phi_1\rangle\langle R,\phi_2\rangle v_{a_1}\dots v_{a_n}\quad \textrm{where } v_{a_i} = \langle Post_{a_i}(R)^\nearrow,\coal S(\phi_1\until\phi_2)\rangle \\
            & \mathcal E(v)(W_{\phi_1}, W_{\phi_2}, W_{a_1},\dotsc,W_{a_n}) = \mathit{Forceable}_S(W_{\phi_1},W_{\phi_2}, \bigcup_i W_{a_i})
        \end{align*}

    \item Case $v=\langle R,\coalbr S(\phi_1\until\phi_2)\rangle$:
        \begin{align*}
            & E(v)=\langle R,\phi_1\rangle\langle R,\phi_2\rangle v_{a_1}\dots v_{a_n}\quad\textrm{where } v_{a_i} = \langle Post_{a_i}(R)^\nearrow,\coalbr S(\phi_1\until\phi_2)\rangle \\
            & \mathcal E(v)(W_{\phi_1}, W_{\phi_2}, W_{a_1},\dotsc,W_{a_n}) = \mathit{Unavoidable}_S(W_{\phi_1},W_{\phi_2}, \bigcup_i W_{a_i})
        \end{align*}
\end{itemize}

\begin{figure}
    \centering
    \begin{tikzpicture}[
                node distance=0.8,
                input/.style = {draw,circle,minimum size=1mm,inner sep=0pt, outer sep=0pt},
                efunc/.style = {inner sep=0.25pt},
                dep/.style = {-stealth}
            ]

        \node[draw] (A_phi) {$\mathbf A,x\leq 4,\phi$};
        \node[efunc, below=0 of A_phi] (A_phi_e) {\tiny$\mathit{Forceable}_{\{\ExamPlayerI,\ExamPlayerIII\}}(\;\;,\;\;,\;\;)$};
        \node[input] (A_phi_1) at ($(A_phi.south) + (5.1mm,-1mm)$) {};
        \node[input] (A_phi_2) at ($(A_phi.south) + (7.6mm,-1mm)$) {};
        \node[input] (A_phi_3) at ($(A_phi.south) + (10.1mm,-1mm)$) {};
        \node[draw,circle,inner sep=1pt,below right=0.5 and 0 of A_phi_3] (A_phi_union) {$\bigcup$};

        \node[draw, below left=0.5 and -0.95 of A_phi] (A_c) {$\mathbf A,x\leq 4,\neg\mathbf C$};
        \node[efunc, below=0 of A_c] (A_c_e) {\tiny$\coalbr{x\leq 4}$};

        \node[draw, below=0.5 of A_c] (A_goal) {$\mathbf A,x\leq 4,\mathbf{Goal}$};
        \node[efunc, below=0 of A_goal] (A_goal_e) {\tiny$\emptyset$};

        \node[draw,right=2.5 of A_phi] (B_phi) {$\mathbf B,x\leq 5,\phi$};
        \node[efunc, below=0 of B_phi] (B_phi_e) {\tiny$\mathit{Forceable}_{\{\ExamPlayerI,\ExamPlayerIII\}}(\;\;,\;\;,\;\;)$};
        \node[input] (B_phi_1) at ($(B_phi.south) + (5.1mm,-1mm)$) {};
        \node[input] (B_phi_2) at ($(B_phi.south) + (7.6mm,-1mm)$) {};
        \node[input] (B_phi_3) at ($(B_phi.south) + (10.1mm,-1mm)$) {};
        \node[draw,circle,inner sep=1pt,below right=0.5 and 0 of B_phi_3] (B_phi_union) {$\bigcup$};

        \node[draw, below left=0.5 and -0.95 of B_phi] (B_c) {$\mathbf B,x\leq 5,\neg\mathbf C$};
        \node[efunc, below=0 of B_c] (B_c_e) {\tiny$\coalbr{x\leq 5}$};

        \node[draw, below=0.5 of B_c] (B_goal) {$\mathbf B,x\leq 5,\mathbf{Goal}$};
        \node[efunc, below=0 of B_goal] (B_goal_e) {\tiny$\emptyset$};

        \node[right=2 of B_phi] (C_phi) {$\cdots$};
        \node[below=0.5 of C_phi] (Goal_phi) {$\cdots$};
        \node[below=0.5 of Goal_phi] (D_phi) {$\cdots$};

        \node[left=of A_phi] (v0) {};

        \draw[dep] (v0) -- (A_phi);
        \draw[dep] (A_phi_1) edge [out=270, in=0, looseness=1.5] (A_c);
        \draw[dep] (A_phi_2) edge [out=270, in=0, looseness=1.5] (A_goal);
        \draw[dep] (A_phi_3) edge [out=270, in=115, looseness=1.0] (A_phi_union);
        \draw[dep] (A_phi_union) edge [out=65, in=180, looseness=1.5] node[above] {$a_1$} (B_phi);
        \draw[dep] (B_phi_1) edge [out=270, in=0, looseness=1.5] (B_c);
        \draw[dep] (B_phi_2) edge [out=270, in=0, looseness=1.5] (B_goal);
        \draw[dep] (B_phi_3) edge [out=270, in=115, looseness=1.0] (B_phi_union);
        \draw[dep] (B_phi_union) edge [out=65, in=180, looseness=1.5] node[above] {$a_2$} (C_phi);
        \draw[dep] (B_phi_union) edge [out=0, in=180, looseness=1.5] node[above] {$a_3$} (Goal_phi);
        \draw[dep] (B_phi_union) edge [out=-65, in=180, looseness=1.0] node[above] {$a_4$} (D_phi);
        
    \end{tikzpicture}
    \caption{A fragment of the EADG for $\phi=\coal{\ExamPlayerI,\ExamPlayerIII}(\neg\mathbf{C}\;\until\;\mathbf{Goal})$ in $\mathcal A$ from \cref{fig:intro_example}.}
    \label{fig:encoding_example}
\end{figure}

We remark that the elements of the NOR $\langle Fed(Q),\subseteq,\emptyset\rangle$ have finite representations using the data structure known as a \emph{difference bound matrix}~\cite{Bengtsson2004TASemanticsAlgoTools,Clarke2018Handbook}. All value functions are effectively computable since all constituent operations have known algorithms on difference bound matrices~\cite{Bengtsson2004TASemanticsAlgoTools,Cassez2005TimedGames}.

In \cref{fig:encoding_example}, we show a fragment of the dependency graph generated when checking if $\phi=\coal{\ExamPlayerI,\ExamPlayerIII}(\neg\mathbf{C}\;\until\;\mathbf{Goal})$ holds in the initial state of $\mathcal A$ from \cref{fig:intro_example}. Eventually, the algorithm assigns $\coalbr{x\leq 3}$ to the vertex $\langle\mathbf A,x\leq 4,\phi\rangle$ and then we can terminate as we now know that the $\phi$ holds for the initial state $\langle\mathbf A,\veco\rangle$.

We will now state the correctness of the encoding and its supporting lemma.

\begin{lemmarep}[Monotonically Safe]\label{lem:monotonically_safe}
    The EADG encoding of TATL has no cycles of non-monotonic value functions and $dist(\langle R,\phi\rangle)$ is the greatest number of nested negations in $\phi$.
\end{lemmarep}
\begin{proof}
    Of all used set operations, the operations that are not monotonically increasing are subtraction, complement, and $Pred_\lambda$. Let us consider their uses:
    \begin{itemize}
        \item For vertex $v=\langle R,\neg\phi\rangle$ we have $\mathcal E(v)(W)=R\setminus W$ which is not monotonically increasing. However, $E(v)=\langle R,\phi\rangle$, and there is no way to have $\langle R,\phi\rangle\to^+\langle R,\neg\phi\rangle$ as all vertices $\langle R', \phi'\rangle$ only depend on vertices $\langle R'',\phi''\rangle$ where $\phi'=\phi''$ or $\phi''$ is a sub-formula of $\phi'$.
        \item The monotonicity of safe timed predecessors $Pred_\lambda (W, W')$ depends on its inputs. It is monotonically increasing if its first parameter $W$ increases monotonically and its second parameter $W'$ decreases monotonically. $Pred_\lambda$ is used in both $\mathit{Forceable}$ and $\mathit{Unavoidable}$ which are only used with monotonically increasing inputs:
        \begin{itemize}
            \item In $\mathit{Forceable}$ (\ref{eq:forceable}), the second parameter to $Pred_\lambda$ is the expression $\overline{W_{\phi_1}}\cup Pred_{\overline S}(\overline{\mathcal W})$ minus $W_{\phi_2}$. Due to the complements, this expression decreases monotonically even though the inputs are monotonically increasing, so this use of $Pred_\lambda$ is monotonically increasing.
            \item In $\mathit{Unavoidable}$ (\ref{eq:avoidable}), the second parameter to $Pred_\lambda$ is the expression $\overline{W_{\phi_1}}\cup (Pred_S(\overline{\mathcal W})\setminus Pred_{\overline{S}}(\mathcal W))$ minus $W_{\phi_2}$ instead. Again, the complements ensure this expression is monotonically decreasing, so this use of $Pred_\lambda$ is also monotonically increasing.
        \end{itemize}
    \end{itemize}
    All other value functions are clearly monotonic and thus we have that the EADG does not contain cycles with non-monotonic functions. Moreover, since only formulae with negations have non-monotonic value functions, the value $dist(\langle R,\phi\rangle)$ is equal to the greatest number of nested negations in $\phi$.
\end{proof}

\begin{theoremrep}[Encoding Correctness]\label{theo:encoding_correctness}
    Given a TMG $\mathcal A=\langle L, \ell_\mathrm{init}, X, A, T, I\rangle$, a set of states $R\in Zone(Q)$ such that $R=R^\nearrow$, a state $\langle\ell,\nu\rangle\in R$, and a TATL formula $\phi$, then $\langle\ell,\nu\rangle\vDash\phi$ iff $\langle\ell,\nu\rangle\in\alpha_{\min}^G(\langle R,\phi\rangle)$ using the presented EADG encoding of TATL.
\end{theoremrep}
\begin{proof}
    To prove \cref{theo:encoding_correctness}, we need the following lemma that gives a lower bound guarantee on the content of dependencies.
    \begin{lemma}\label{lem:post_a_delay}
        For all generated vertices $\langle R,\phi\rangle\in V$, we have that if $\langle\ell,\nu\rangle\in R$ and $\langle\ell,\nu\rangle\xto\delta\langle\ell,\nu'\rangle$ for some $\delta\geq0$ then $\langle\ell,\nu'\rangle\in R$. That is $R=R^\nearrow$.
    \end{lemma}

    \cref{lem:post_a_delay} follows from our encoding where all $\langle R,\phi\rangle\in V$ depend only on vertices $\langle R',\phi'\rangle$ where $R'=R$ or $R'=R'^\nearrow$, and the initial vertex $\langle\ell_\textrm{init},\{\veco\}^\nearrow\cap\coalbr{I(\ell_\textrm{init})}, \phi'\rangle$ is also closed under delays.

    We now return to proving \cref{theo:encoding_correctness}.
    Given TATL property $\phi$, we denote by $\coalbr{\phi}=\{q\in Q\mid q\vDash\phi\}$ all states that satisfy $\phi$. The theorem's claim can then be restated as $\alpha_{\min}^G(\langle R,\phi\rangle)=R\cap\coalbr{\phi}$.
    By \cref{lem:monotonically_safe} on monotonicity, we only need to argue that if $E(v)=v_1\cdots v_n$ then $\mathcal E(v)(\alpha_{\min}^G(v_1),\dotsc,\alpha_{\min}^G(v_n))=\alpha_{\min}^G(v)$. That is, the value function at vertex $v$ evaluates to the fixed-point assignment of $v$ if we input the fixed-point assignments of the dependencies $E(v)$.
    The proof now follows from structural induction on vertex $v\in V$.
    \begin{itemize}
        \item Let $v=\langle R,\pi\rangle$ with $\pi\in\Pi$: Vertex $v$ has no dependencies and $\alpha_{\min} ^G(v)=\mathcal E(v)(\varepsilon)=\{\langle\ell,\nu\rangle\in R\mid \pi\in Lab(\ell)\}= R\cap\coalbr{\pi}$.
        \item Let $v=\langle R,g\rangle$ with $g\in C(X)$: Vertex $v$ has no dependencies and $\alpha_{\min} ^G(v)=\mathcal E(v)(\varepsilon)=\{\langle\ell,\nu\rangle\in R\mid \nu\in\coalbr{g}\}= R\cap\coalbr{g}$.
        \item Let $v=\langle R,\neg\phi\rangle$: We have $E(v)=\langle R,\phi\rangle$ and by induction hypothesis, $\alpha_{\min}^G(\langle R,\phi\rangle)=R\cap\coalbr{\phi}$. It follows that $\alpha_{\min}^G(v)=\mathcal E(v)(R\cap\coalbr{\phi})=R\setminus (R\cap\coalbr{\phi})=R\cap\overline{\coalbr{\phi}}=R\cap\coalbr{\neg\phi}$.
        \item Let $v=\langle R,\phi_1\lor\phi_2\rangle$: We have $E(v)=\langle R,\phi_1\rangle\langle R,\phi_2\rangle$. By induction hypothesis, $\alpha_{\min}^G(\langle R,\phi_1\rangle)=R\cap\coalbr{\phi_1}$ and $\alpha_{\min}^G(\langle R,\phi_2\rangle)=R\cap\coalbr{\phi_2}$. It follows that $\alpha_{\min}^G(v)=\mathcal E(v)(R\cap\coalbr{\phi_1}, R\cap\coalbr{\phi_2})=(R\cap\coalbr{\phi_1})\cup(R\cap\coalbr{\phi_2})=R\cap(\coalbr{\phi_1}\cup\coalbr{\phi_2})=R\cap\coalbr{\phi_1\lor\phi_2}$.
        \item Let $v=\langle R,\phi_1\land\phi_2\rangle$: We have $E(v)=\langle R,\phi_1\rangle\langle R,\phi_2\rangle$. By induction hypothesis, $\alpha_{\min}^G(\langle R,\phi_1\rangle)=R\cap\coalbr{\phi_1}$ and $\alpha_{\min}^G(\langle R,\phi_2\rangle)=R\cap\coalbr{\phi_2}$. It follows that $\alpha_{\min}^G(v)=\mathcal E(v)(R\cap\coalbr{\phi_1}, R\cap\coalbr{\phi_2})=(R\cap\coalbr{\phi_1})\cap(R\cap\coalbr{\phi_2})=R\cap\coalbr{\phi_1\land\phi_2}$.
        \item Let $v=\langle R,z.\phi\rangle$ with $z\in X$: We have $E(v)=\langle R[z]^\nearrow,\phi\rangle$ where $R[z]^\nearrow$ contains all states reachable from a state in $R$ by some delay given the formula clock $z$ is reset first. By induction, $\alpha_{\min}^G(\langle R[z]^\nearrow,\phi\rangle)=R[z]^\nearrow\cap\coalbr{\phi}$. It follows that $\alpha_{\min}^G(v)=\mathcal E(v)(R[z]^\nearrow\cap\coalbr{\phi})=(R[z]^\nearrow\cap\coalbr{\phi}\cap Q[z])\#z\cap R$. The set $\coalbr{\phi}\cap Q[z]$ contains all states where $\phi$ holds and $z$ is 0. The free operation turns the set into all states where $\phi$ holds if $z$ is reset, i.e.\ $(\coalbr{\phi}\cap Q[z])\#z=\coalbr{z.\phi}$. The intersections with $R[z]^\nearrow$ and $R$ reduce the set to what we know and care about, and we get $\alpha_{\min}^G(v)=R\cap\coalbr{z.\phi}$.
        \item Let $v=\langle R,\coal{S}\bigcirc\phi\rangle$: The dependencies are vertices $\langle Post_{a_i}(R)^\nearrow,\phi\rangle$ in a fixed order by $a_i\in A$. By induction hypothesis the input to the value function is a string of sets $W_{a_i}=Post_{a_i}(R)^\nearrow\cap\coalbr{\phi}$ similarly ordered, and $\mathcal E(v)(W_{a_1},\dotsc,W_{a_n}) = \mathit{Forceable}_S(R,\emptyset,\bigcup_i W_{a_i})$. Since the 2nd input is $\emptyset$, this set only contains states that can delay to a state with a discrete action to some $W_{a_i}$, and since the 1st input is $R$, we put no further conditions on which states can be delayed in. Lastly, since $W_{a_i}$ contains all states that satisfy $\phi$ and are reachable from $R$ through a delay followed by action $a_i$ by \cref{lem:post_a_delay}, we get (by construction) that $\alpha_{\min}^G(v)=R\cap\coalbr{\coal S\bigcirc\phi}$.
        \item Let $v=\langle R,\coal{S}(\phi_1\;\until\;\phi_2)\rangle$: The dependencies are $\langle R,\phi_1\rangle$, $\langle R,\phi_2\rangle$, and a string of $\langle Post_{a_i}(R)^\nearrow,\coal{S}(\phi_1\;\until\;\phi_2)\rangle$ in a fixed order by $a_i\in A$. By induction hypothesis the inputs to the value function are $W_{\phi_1}=R\cap\coalbr{\phi_1}$, $W_{\phi_2}=R\cap\coalbr{\phi_2}$, and a string of sets $W_{a_i}=Post_{a_i}(R)^\nearrow\cap\coalbr{\phi}$ ordered by $a_i$, and we have $\mathcal E(v)(W_{\phi_1},W_{\phi_2},W_{a_1},\dotsc,W_{a_n}) = \mathit{Forceable}_S(W_{\phi_1},W_{\phi_2},\bigcup_i W_{a_i})$. Since the 1st input is $W_{\phi_1}$ and the second input is $W_{\phi_2}$, we get that $\phi_1$ must hold unless (or until) $\phi_2$ holds. By \cref{lem:post_a_delay} and since the 2nd input is $W_{\phi_2}$, the output will contain states where coalition $S$ can safely delay into a state satisfying $\phi_2$. Lastly, since $W_{a_i}$ contains states that satisfy $\coal{S}(\phi_1\;\until\;\phi_2)$ and, by \cref{lem:post_a_delay}, are reachable from $R$ through a delay followed by action $a_i$, we get (by construction) that $\alpha_{\min}^G(v)=R\cap\coalbr{\coal{S}(\phi_1\;\until\;\phi_2)}$.
        \item Let $v=\langle R,\coalbr{S}(\phi_1\;\until\;\phi_2)\rangle$: Again the dependencies are $\langle R,\phi_1\rangle$, $\langle R,\phi_2\rangle$, and a string of $\langle Post_{a_i}(R)^\nearrow,\coalbr{S}(\phi_1\;\until\;\phi_2)\rangle$ in a fixed order by $a_i\in A$. By induction hypothesis the inputs to the value function are $W_{\phi_1}=R\cap\coalbr{\phi_1}$, $W_{\phi_2}=R\cap\coalbr{\phi_2}$, and a string of sets $W_{a_i}=Post_{a_i}(R)^\nearrow\cap\coalbr{\phi}$ ordered by $a_i$, and we have $\mathcal E(v)(W_{\phi_1},W_{\phi_2},W_{a_1},\dotsc,W_{a_n}) = \mathit{Unavoidable}_S(W_{\phi_1},W_{\phi_2},\bigcup_i W_{a_i})$. Since the 1st input is $W_{\phi_1}$ and the second input is $W_{\phi_2}$, we get that $\phi_1$ must hold unless (or until) $\phi_2$ holds. Moreover, by \cref{lem:post_a_delay}, the output will contain states where the opposition can delay into a state satisfying $\phi_2$ unless $S$ takes a discrete action to outside all $W_{a_i}$. Lastly, since $W_{a_i}$ contains states that satisfy $\coalbr{S}(\phi_1\;\until\;\phi_2)$ and, by \cref{lem:post_a_delay}, are reachable from $R$ through a delay followed by action $a_i$, we get (by construction) that $\alpha_{\min}^G(v)=R\cap\coalbr{\coalbr{S}(\phi_1\;\until\;\phi_2)}$.
    \end{itemize}
\end{proof}

\subsection{Unsatisfied states}\label{sec:unsat_nor}

We now detail an improvement also investigated in \cite{Cassez2005TimedGames}.
In the encoding presented above, the NOR domain represents the states for which the formula is definitely satisfied. It is also possible to define an encoding in which the NOR domain represents the states in which the formula is definitely not satisfied. It is easy to see that the composition of two NORs is also a NOR~\cite{Enevoldsen2019ADGs}, so we can do both simultaneously. If the initial state is ever included in the set of unsatisfied valuations  then we know that the property does not hold. Hence, with this extension, we can terminate early for many negative cases as well.
The encoding using the unsat NOR is trivial for clock constraints $g\in C(X)$, and operators $\land$, $\lor$, and $\neg$, but for completeness, we state the encoding of formulae containing the $\bigcirc$ and $\until$ operator below. For all $v\in V$, we use the same edges $E(v)$, but different value functions, denoted $\mathcal E^c$, for updating the unsatisfied part of the composed NOR. Here $M$ is used to denote elements in the unsat NOR and we assume the same ordering of actions $A=\{a_1,\dotsc,a_n\}$ from earlier:

\begin{itemize}
    \item Case $v=\langle R,\coal S\bigcirc \phi$:
    $$
        \mathcal E^c(v)(M_{a_1},\dotsc,M_{a_n})=\mathit{Unavoidable}_S(R,\emptyset, \bigcup_i M_{a_i})
    $$
    
    \item Case $v=\langle R,\coal S(\phi_1\until\phi_2)$:
    $$
        \mathcal E^c(v)(M_{\phi_1},M_{\phi_2},M_{a_1},\dotsc,M_{a_n})=\mathit{Unavoidable}_S(M_{\phi_2}, M_{\phi_1}, \bigcup_i M_{a_i})
    $$
    
    \item Case $v=\langle R,\coalbr S(\phi_1\until\phi_2)$:
    $$
        \mathcal E^c(v)(M_{\phi_1},M_{\phi_2},M_{a_1},\dotsc,M_{a_n})=\mathit{Forceable}_S(M_{\phi_2}, M_{\phi_1}, \bigcup_i M_{a_i})
    $$
\end{itemize}

As seen, the difference between $\mathcal E$ and $\mathcal E^c$, is that every use of $\mathit{Forceable}$ has been replaced by $\mathit{Unavoidable}$, and vice versa, and the terms involving $\phi_1$ and $\phi_2$ have been swapped in $\until$ formulae.

\section{Dynamical Vertex Merge in EADGs}\label{sec:vertex_merging}

Usually, automated verification of timed systems takes advantage of inclusion checking, i.e.\ if one discovered symbolic state is included within another, then there is no reason to investigate the smaller one. To do this in the EADG framework, we must first introduce the general concept of vertex merging, which can be used to dynamically reduce the size of the graph.

Consider the NOR $\langle\mathcal D,\sqsubseteq,\bot\rangle$ and an EADG $G=\langle V,E,\mathcal D,\mathcal E\rangle$. If $u\in V^*$ is a string of vertices, then $u[i]$ denotes the $i$th vertex in the string and $u[v_1\mapsto v_2]$ denotes the string $u$ where all occurrences of $v_1$ are replaced with $v_2$. Given an $n$-ary function $h: \mathcal D^n \rightarrow \mathcal D$, a unary function $f: \mathcal D \rightarrow \mathcal D$ and an index $i$, $1 \leq i \leq n$, we use $\circ_i$ to denote pointwise function composition on the $i$th input, defined as $(h\circ_i f)(x_1,\dotsc,x_n)=h(x_1,\dotsc,f(x_i),\dotsc,x_n)$, and its extension to multiple points $\circ_{\{i_1,\dotsc,i_k\}}$ given by $(h\circ_{\{i_1,\dotsc,i_k\}} f)=((h\circ_{i_1} f) \circ_{i_2} \cdots) \circ_{i_k} f$ where notably $h\circ_\emptyset f=h$.

\begin{definition}[Derivation]
    Let $v_1,v_2\in V$ be vertices in EADG $G=\langle V,E,\mathcal D,\mathcal E\rangle$ with $dist(v_1)\leq dist(v_2)$ and let $f:\mathcal D\to \mathcal D$ be a monotonic function such that $f(\alpha_{\min}^G(v_2))=\alpha_{\min}^G(v_1)$. We call $f$ a \emph{derive function} and say that $v_1$ is \emph{derivable} from $v_2$ through $f$, denoted by $v_1\preceq_f v_2$.
\end{definition}

Derivable vertices can be removed by merging without loss of precision.

\begin{definition}[Vertex Merge]\label{def:vertex_merge}
    Let $v_1,v_2\in V$ be vertices in EADG $G=\langle V,E,\mathcal D,\mathcal E\rangle$ with $dist(v_1)\leq dist(v_2)$ and let $f:\mathcal D\to\mathcal D$ be a monotonic function such that $v_1\preceq_f v_2$. A \emph{vertex merge of $v_1$ into $v_2$ by $f$} is an operation that results in a new EADG denoted $G[v_1\mapsto_f v_2]=\langle V',E',\mathcal D,\mathcal E'\rangle$ where
    \begin{itemize}
        \item $V'=V\setminus\{v_1\}$,
        \item $E'(v)=E(v)[v_1\mapsto v_2]$ for all $v\in V'$, and
        \item $\mathcal E'(v)=\mathcal E(v)\circ_I f$ for all $v\in V'$ where $I=\{i\mid E(v)[i]=v_1\}$.
    \end{itemize}
\end{definition}

We can now present a theorem that shows that a vertex merge does not change the minimum fixed-point value of any vertex in the dependency graph.

\begin{theoremrep}[Merge Preserves $\alpha_{\min}^G$]\label{theo:merge_preserves_fixed_point}
    Let $G=\langle V,E,\mathcal D,\mathcal E\rangle$ be an EADG. If $v_1,v_2\in V$ and $f:\mathcal D\to\mathcal D$ such that $v_1\preceq_f v_2$ then $\alpha_{\min}^G(v)=\alpha_{\min}^{G[v_1\mapsto_f v_2]}(v)$ for all $v\in V\setminus\{v_1\}$.
\end{theoremrep}
\begin{proof}
    Since merges require $dist(v_1)\leq dist(v_2)$ and that $f$ is monotonic, they cannot introduce new non-monotonic cycles in the graph.
    Moreover, only $v\to v_1$, the direct dependents of $v_1$, are structurally affected by a merge. That is, every occurrence of $v_1$ in $E(v)$ is replaced by $v_2$, and $\mathcal E(v)$ is pointwise composed with $f$ at indices that were previously $v_1$. Since $f(\alpha_{\min}^G(v_2))=\alpha_{\min}^G(v_1)$ by definition, the updated $\mathcal E'(v)$ receives the same values in the context of $\alpha_{\min}^G$, and due to the preserved monotonicity, this is enough to conclude the theorem.
\end{proof}

There are two main reasons why merging vertices can be beneficial when $v_1\preceq_f v_2$. First, $f$ may be computationally cheaper than computing $\mathcal E(v_1)$. Second, it reduces the size of the graph, making other operations such as back-propagation of updates and pruning cheaper.
However, if $f$ is expensive, it may be better to keep $v_1$ for the memoization it provides.

On \cref{fig:eadge_example:post_merge}, we show how the EADG from \cref{fig:eadge_example:pre_merge} looks after a merge by derivation $E\preceq_f F$ where $f(x)=x+3$. The existence of this derivation depends on the user's domain knowledge. In this case, the user might know that the first input to the max function of $E$ can never exceed 3, and hence its value can be derived from the second input $F$. As shown in the figure, the resulting graph is smaller and it is also possible to prune the vertex $H$.

\subsection{Algorithm with Vertex Merging}\label{app:algorithm}

\SetKwFor{Proc}{proc}{:}{end}
\begin{algorithm}
    \caption{Minimum fixed-point computation on an EADG. The lines highlighted in gray are our additions to the algorithm from \cite{Enevoldsen2022EADGs}.}
    \label{algo:fixed_point_eadg_with_merge}
    \scriptsize
    \KwIn{An EADG $\langle V, E,\mathcal D,\mathcal E\rangle$ and $v_0\in V$}
    \KwOut{$d\in\mathcal D$ s.t. $d\sqsupseteq\alpha_{\min}(v_0)$}
    \BlankLine
    
    $Dep(v)=\emptyset$, $Edges(v):=E(v)$, and $Eval(v)=\mathcal E(v)$ for all $v\in V$\;
    $\hat v:=v_0$; $\hat f=id$; $\alpha:=\alpha_\bot$\;
    $W:=\{v_0\}$; $Pass:=\emptyset$; $Active:=\{v_0\}$\;

    \While{$W\neq\emptyset$}{
        let $v\in W$ where $v$ is \textit{pickable}\;
        $W:=W\setminus\{v\}$\;
        $\textsc{PruneDependents}(v)$\;
        \If{$v=\hat v$ or $Dep(v)\neq\emptyset$}{
            \If{$v\notin Pass$}{
                $v:=\textsc{Explore}(v)$\;
            }
            \If{$v\in Pass$ or $\mathcal E(v)$ is non-monotonic}{
                let $v_1v_2\dots v_k:=Edges(v)$\;
                $d:=\mathcal E(v)(\alpha(v_1),\alpha(v_2),\dotsc,\alpha(v_k))$\;
                \If{$\alpha(v)\sqsubset d$}{
                    $W:=W\cup\{u\in Dep(v)\mid \exists i . Edges(u)[i]=v\land i\notin\textsc{Ignore}(\alpha, u)\}$\;
                    $\alpha(v):=d$\;
                    \If{$v=\hat v$ and $\{1,2,\dotsc,k\}\subseteq\textsc{Ignore}(\alpha, \hat v)$}{
                        \KwRet{$\hat f(\alpha(\hat v))$}
                    }
                }
            }
            $Pass:=Pass\cup\{v\}$\;
        }
    }
    \KwRet{$\hat f(\alpha(\hat v))$}
    \BlankLine

    \Proc{$\textsc{Explore}(v)$}{
        $v_*=v$\;
        \For{$i=1$\ \KwTo $|Edges(v)|$}{
            let $v_i=Edges(v)[i]$ (generate it if needed)\;
            \If{$v_i\notin Active$}{
                $Active:=Active\cup\{v_i\}$\;
                \bhighlight
                \ForAll{$v'\in Active$ \emph{\textbf{where}} $v'\neq v_i$}{
                    \If{$v_i\preceq_f v'$ \emph{\textbf{for some}} $f$}{
                        $Active:=Active\setminus\{v_i\}$\;
                        $Eval(v):=Eval(v)\circ_i f$\;
                        $Edges(v):=Edges(v)[v_i\mapsto v']$\;
                        \textbf{continue outer loop}\;
                    }
                    \If{$v'\preceq_f v_i$ \emph{\textbf{for some}} $f$}{
                        $\textsc{Replace}(v',f,v_i)$\;
                        \If{$v'=\hat v$}{
                            $\hat v:=v_i$\;
                            $\hat f:=\hat f\circ f$\;
                        }
                        \If{$v'= v_*$}{
                            \tcc{We replaced the vertex we were exploring. Record replacement.}
                            $v_*:=v_i$\;
                        }
                    }
                }
            }
            \If{$v_*\neq v$}{
                \KwRet{$\textsc{Explore}(v_i)$}\ehighlight{lightgray}
            }
            $Dep(v_i):=Dep(v_i)\cup\{v\}$\;
            \lIf{$v_i\notin Pass$}{$W:=W\cup\{v_i\}$}
        }
        \KwRet{$v$}
    }
    \BlankLine
    \bhighlight
    \Proc{$\textsc{Replace}(v_1, f, v_2)$}{
        $Dep(v_2):=Dep(v_1)$\;
        \ForAll{$v'\in Active\cap Dep(v_1)$}{
            $Eval(v'):=Eval(v')\circ_I f$ where $I=\{i\mid Edges(v')[i]=v_1\}$\;
            $Edges(v'):=Edges(v')[v_1\mapsto v_2]$
        }
        $Pass:=Pass\setminus\{v_1\}$\;
        $Active:=Active\setminus\{v_1\}$\;
        $W:=W\setminus\{v_1\}$\ehighlight{lightgray}\;
    }
    \BlankLine
    \Proc{$\textsc{PruneDependents}(v)$}{
        $C:=\{u\in Dep(v)\mid \forall i . Edges(u)[i]=v\implies i\in\textsc{Ignore}(\alpha, u)\}$\;
        $Dep(v):=Dep(v)\setminus C$\;
        \If{$Dep(v)=\emptyset$ and $C\neq\emptyset$}{
            $Pass:=Pass\setminus\{v\}$\;
            $\textsc{UpdateDependentsRec}(v)$\;
        }
    }
    \BlankLine
    \Proc{$\textsc{PruneDependentsRec}(v)$}{
        \ForAll{$u\in E(v)$}{
            $C:=Dep(u)\cap v$\;
            $Dep(u):=Dep(u)\setminus\{v\}$\;
            \If{$Dep(u)=\emptyset$ and $C\neq\emptyset$}{
                $\textsc{PruneDependentsRec}(u)$\;
                $Pass:=Pass\setminus\{u\}$\;
            }
        }
    }
\end{algorithm}

Here we present our modification to the fixed-point algorithm from~\cite{Enevoldsen2022EADGs}, enabling it to take advantage of vertex merging whenever $v_1\preceq_f v_2$. The updated algorithm uses the following data structures:
\begin{itemize}
    \item $\hat v$ is the current root vertex, initialized to $v_0$,
    \item $\hat f$ is the root deriving function, initialized to $id$ (the identity function),
    \item $\alpha:V\to \mathcal D$ is the current assignment, initialized to $\alpha_\bot$,
    \item $W$ is a waiting set of vertices pending exploration or reevaluation,
    \item $Pass$ is a set of explored vertices,
    \item $Dep:V\to 2^V$ is a function that for each vertex $v$ returns a set of dependent vertices that should be reevaluated if the assignment of $v$ changes,
    \item $Edges:V\to V^*$ is a function that for each vertex $v$ returns the current edges of $v$, initialized to $E(v)$,
    \item $Eval(v):\mathcal D^n\to\mathcal D$ is the current value function for vertex $v$ where $n=|E(v)|$, initialized to $\mathcal E(v)$,
    \item $Active$ is a set of vertices that are (still) relevant for the root node $\hat v$ and considered during vertex merging.
\end{itemize}
The variables $\hat v$ and $\hat f$, and the data structures $Edges$, $Eval(v)$, and $Active$ are introduced by us to keep track of the reduced graph. In \cref{algo:fixed_point_eadg_with_merge}, we show the EADG minimum fixed point computation algorithm from \cite{Enevoldsen2022EADGs} updated with vertex merging.\footnote{We also rearranged and renamed some procedures for clarity. Additionally, \textsc{Ignore} returns a set of indices instead of vertices in our version.} 
Incorporating vertex merging is relatively straightforward, but there are some edge cases to discuss.
Vertex merging, highlighted in gray in \cref{algo:fixed_point_eadg_with_merge}, occurs during the exploration of the vertex $v$ (line 21-44). Each dependency $v_i$ of $v$ is checked against all other active vertices. If $v_i$ can be merged into another active vertex $v'$, then we make $v$ depend on $v'$ instead and make $v_i$ inactive (lines 28-32). If an active vertex $v'$ can be merged into $v_i$ then we make $v'$ inactive and replace $v'$ with $v_i$ in all our data structures (line 34). Additionally, if $v'$ is the root vertex $\hat v$ then we make $v_i$ the new root and update $\hat f$ by composing it with $f$ so that the root assignment can be derived later (lines 35-37). In (presumably) rare cases, $v'$ is the explored vertex $v$, and if so, exploration must be restarted (lines 38-41).

\begin{lemmarep}\label{lem:explore_procedure}
    At the start and end of the $\textsc{Explore}$ procedure, there exists no $v_1,v_2\in Active$ such that $v_1\neq v_2$ and $v_1\preceq_f v_2$.
\end{lemmarep}
\begin{proof}
    First, we note that $Active$ initially only contains $v_0$ and that it is only modified in the $\textsc{Explore}$ procedure, meaning the lemma holds on the first entry.
    Now let us consider the $\textsc{Explore}$ procedure in depth.
    On line 22, we record the original vertex $v$ as $v_*$ since we might merge it away later.
    The successors of the input vertex $v$ are established on lines 23-24. It is important for correctness that the successors are found/generated one by one. Some of the vertices may already exist while others have to be generated. If successor $v_i$ is not active already, it must be new. We mark new successors as active on line 26 and then compare it to all other active vertices in the loop on line 27 and forward. By hypothesis, no pair of active vertices can be merged when the algorithm enters the $\textsc{Explore}$ procedure, so with the introduction of successor $v_i$, only one of the following three cases can apply:
    \begin{enumerate}
        \item $v_i$ can be merged into one or more active vertices $v'$,
        \item one or more active vertices $v'$ can be merged into $v_i$,
        \item $v_i$ cannot merge with any active vertex.
    \end{enumerate}
    Case 1 is handled on lines 28-32. We set $v_i$ to inactive, compose $f$ onto $Eval(v)$ on the $i$th index, and replace all occurrences of $v_i$ with $v'$ in $Edges(v)$. These updates correspond to a merge operation as described in \cref{def:vertex_merge}, and since $v_i$ is a newly generated vertex, $v$ is its only dependent. Of course, now that $v_i$ has been merged into $v'$, it will not be a part of the graph and cannot be merged into anything else either, and hence we continue to the next iteration of the outer loop.
    Case 2 is handled on lines 33-39. First, we call $\textsc{Replace}(v',f,v_i)$. This procedure will replace all usages of $v'$ with $v_i$, applying the aforementioned merge operations, but it also updates $Dep$, which keeps track of dependents. The dependents of $v'$ are simply added as dependents of $v_i$. Next, we need to handle some edge cases. On lines 35-37, if $v'=\hat v$, then we just replaced the current root node and we need to update $\hat v$ to point to $v_i$ and compose $f$ onto the root derive function $\hat f$. On lines 38-39, if $v'=v_*$, then we just replaced the vertex we are currently exploring, and further exploration of that node is meaningless. However, we must first check if there are more active nodes that can be replaced by $v_i$, so just record $v_i$ as $v_*$ for now. Once the loop is done, we explore the replacement $v_*$ of $v$ on line 41. On arrival at line 40, all three cases from before have been checked and the invariant from the lemma now holds again. Hence, no matter if we exit on line 41 or proceed to handle zero or more successors, our invariant holds, concluding our proof.
\end{proof}

\begin{theoremrep}
    The algorithm terminates and returns $\alpha_{\min}^G(v_0)$.
\end{theoremrep}
\begin{proof}
    For this proof, we rely on many of the lemmas from~\cite{Enevoldsen2022EADGs}. In fact, we primarily need to argue that the graph induced by the content of data structures $Edges$, $Eval$, and $Active$ is the reduced graph that results of applying vertex merges, and that $\hat v$ and $\hat f$ can be used to derive the assignment of the original $v_0$.

    By \cref{lem:explore_procedure}, we know that no merges are possible at the end of the $\textsc{Explore}$ procedure. As argued in the proof for that lemma, this follows from the fact that we apply all possible merges every time a new vertex is discovered. Each merge updates the three data structures $Edges$, $Eval$, and $Active$, so by the end of $\textsc{Explore}$ they contain our reduced graph. Only a finite number of merges are possible, so our addition does not affect termination. By \cref{theo:merge_preserves_fixed_point}, the merges preserve the minimum fixed-point assignment of the remaining vertices.

    As also touched upon in the proof of \cref{lem:explore_procedure}, we update the $Dep$ data structure when we merge away an existing vertex. $Dep(v)$ holds a set of vertices that need to be updated when the assignment of $v$ changes. Upon a merge based on $v'\preceq_f v_i$, the set $Dep(v_i)$ inherits the vertices of $Dep(v')$, and hence the vertices will still be correctly updated.

    Lastly, we also update $\hat v$ and $\hat f$ on lines 36-37 when we merge away the previous root $\hat v$. By \cref{def:vertex_merge}, we can derive the assignment of the previous root using $f(\alpha(v_i))$. As the root vertex is replaced, possibly multiple times, the root derive function $\hat f$ becomes a composition of all derive functions that have been involved. Thus, $\hat f(\alpha(\hat v))$ derives the "current" assignment of $v_0$. When no more updates are possible (line 20) or the assignment of $\hat v$ stabilizes (line 18), we return $\hat f(\alpha(\hat v))$ which is at that point equivalent to $\alpha_{\min}^G(v_0)$.
\end{proof}

\subsection{Inclusion Checking as Derivation}

Finally, we can do vertex merging in our problem domain. If one symbolic state is included within another and the associated query is the same, we merge the smaller one into the bigger one, since it contains the same valuations and more (conventional inclusion checking).
\begin{theoremrep}[Derivation by Intersection]\label{theo:merge_by_intersection_validity}
    Let $\langle R,\phi\rangle,\langle R',\phi\rangle\in V$ be vertices and let $f_R:\mathcal D\to\mathcal D$ be a function such that $f_R(W)=W\cap R$. If $R\subseteq R'$, then $\langle R,\phi\rangle\preceq_{f_R}\langle R',\phi\rangle$ is a derivation.
\end{theoremrep}
\begin{proof}
    By \cref{lem:monotonically_safe}, we know that $dist(\langle R,\phi\rangle=dist(\langle R',\phi\rangle)$.
    By \cref{theo:encoding_correctness}, we have $\alpha_{\min}^G(\langle R,\phi\rangle)=R\cap\coalbr{\phi}$ and $\alpha_{\min}^G(\langle R',\phi\rangle)=R'\cap\coalbr{\phi}$. Since $R\subseteq R'$, clearly $f_R(\alpha_{\min}^G(\langle R',\phi\rangle))=R'\cap\coalbr{\phi}\cap R=R\cap\coalbr{\phi}$, which is exactly what we need.
\end{proof}

\subsection{Expansion Abstraction}\label{sec:expand_abstraction}
It is known that EADGs can be refined using abstractions~\cite{Enevoldsen2022EADGs,Jensen2023DynExtrap,Jensen2025TokenElim}.
We now present a simple abstraction for our domain that has superior performance compared to the inclusion checking.
As discussed, $R$ is a symbolic state and $\langle \ell, Z,\phi\rangle$ is the actual form of vertices in the dependency graph.
For any vertex $v=\langle \ell, Z,\phi\rangle$ there exists a vertex $v'=\langle \ell, \coalbr{I(\ell)},\phi\rangle$ and $v$ can be merged into $v'$ if it is generated. $v'$ will not be merged into another vertex as it is the biggest w.r.t.\ our derivation function in \cref{theo:merge_by_intersection_validity}. Assuming that $v'$ has a high chance of being generated whenever $v$ is, it makes sense to use an abstraction that takes $v$ directly to $v'$ immediately, as the merge is likely to happen later anyway, and their value functions require a similar number of operations but $v'$ holds more information. Doing so for all vertices induces a new smaller EADG where inclusion checking and merging are unnecessary, as for every location there is a unique zone to be explored. In other words, our abstracted vertices are defined only by the discrete part of the state. If the assumption is wrong, i.e.\ $v'$ is not discovered although $v$ was, we may end up exploring states and edges that are not relevant in practice as more discrete transitions may be enabled in the the zone $\coalbr{I(\ell)}$ than in $Z$. However, we assess that timed automata with useless edges are rare in real use cases, and most valuations are eventually reached. As a consequence, the downside of this abstraction is insignificant, as also demonstrated by our experiments.

Formally, we apply the following \emph{expansion abstraction} $\mathfrak X:V\to V$ to each vertex upon generation: 
\begin{align}\label{eq:expand_abstraction}
    \mathfrak X (\langle\ell,Z,\phi\rangle) = \langle\ell,\coalbr{I(\ell)},\phi\rangle \ .
\end{align}
\begin{theoremrep}
    Applying abstraction $\mathfrak X$ from \cref{eq:expand_abstraction} to all vertices preserves encoding correctness, i.e. $\langle\ell,\nu\rangle\vDash\phi$ iff $\langle\ell,\nu\rangle\in\alpha_{\min}^{\mathfrak X(G)}(\langle R,\phi\rangle)$ for every $\langle\ell,\nu\rangle\in R$.
\end{theoremrep}
\begin{proof}
    We sketch two lines of arguments for this theorem. In the first, we establish that effectively $\mathcal E(v)\circ_i f_R=\mathcal E(v)$ for all $v\in V$ and all $i$ such that $E(v)[i]=\langle R,\phi\rangle$, i.e.\ that pointwise composition with the derive function $f_R$ never affects the value function $\mathcal E(v)$ in practice. By construction, if $v\to\langle R,\phi\rangle$ then $R$ contains everything affecting $v$, and for that dependency only the content of $R$ can affect $v$. If vertex $\langle R,\phi\rangle$ is merged into some vertex with $R'\supseteq R$, then $E(v)$ is redirected to use $R'$ instead, but $\mathcal E(v)$ still only uses the subset $R$ anyway. This implies that applying $\mathfrak X$ is effectively equivalent to artificially inserting $\langle\ell,\coalbr{I(\ell)},\phi\rangle$ into the graph and then merging $\langle\ell,Z,\phi\rangle$ into it, and we get the same correctness guarantee. In the second line of argumentation, we build upon the proof for \cref{theo:encoding_correctness} and conclude that the correctness of all value functions $\mathcal E(\langle\ell,Z,\phi\rangle)$ is independent of the shape of $Z$ (as long as the root vertex is closed under the delay operator $\cdot^\nearrow$). Our encoding ensures that $\alpha_{\min}^G(\langle R,\phi\rangle)$ is exactly the subset of $R$ where $\phi$ holds, and thus $R$ is simply a restriction to states that are reachable. If we relax this restriction with the abstraction $\mathfrak X$, we simply find a wider set of states where $\phi$ holds.
    Thus the correctness is preserved by $\mathfrak X$ and $\langle\ell,\nu\rangle\vDash\phi$ iff $\langle\ell,\nu\rangle\in\alpha_{\min}^{\mathfrak X(G)}(\langle R,\phi\rangle)$ when $\langle\ell,\nu\rangle\in R$.
\end{proof}

If follows, that the vertex merging using the derivation we defined in \cref{theo:merge_by_intersection_validity} becomes redundant when using $\mathfrak X$ because there is only one vertex per location-formula pair. The forward exploration done by the algorithm now only considers whether there exists a valuation such that a location can transition another location. This is less strict than a traditional reachability analysis, but it avoids expensive inclusion checks of zones while still restricting the backward propagation to the location-formula pairs we care about.

\section{Implementation and Experimental Evaluation}\label{sec:experiments}

We implement our algorithm in \uppaal~\cite{Hendriks2006Uppaal}. 
We use two stacks for the waiting list, one for vertices waiting for exploration and one for vertices waiting for updates. We always prioritize vertices from the update stack over vertices from the exploration stack, as this was found to be favorable in~\cite{Jensen2023ElimDetached}.
To improve performance, the EADG framework also allows us to specify a function $\textsc{Ignore}(\alpha, v)$ describing which vertices of $E(v)$ can be skipped given the current assignment $\alpha$. We use the universally sound ignore function suggested in \cite{Enevoldsen2022EADGs} by leveraging that $\alpha(\langle R,\phi\rangle) \subseteq R$.
That is: 
\begin{align}
    \textsc{Ignore}(\alpha, \langle R,\phi\rangle)=\begin{cases}
        \N & \textrm{if } \alpha(\langle R,\phi\rangle) = R \\
        \emptyset & \textrm{otherwise}
    \end{cases}
\end{align}
We can reduce the number of non-monotonic functions and thus graph components by rewriting the TATL formulae and pushing negations downward to the atomic propositions when possible.

\paragraph*{The Benchmark}
We benchmark various configurations:
\begin{itemize}
    \item \textsc{Equal}: The traditional EADG algorithm without vertex merging.
    \item \textsc{Incl}: We use vertex merging through inclusion checking as by \cref{theo:merge_by_intersection_validity}.
    \item \textsc{Expand}: We use the expansion abstraction $\mathfrak X$ described in \cref{sec:expand_abstraction} (and no vertex merging).
    \item \textsc{+Unsat}: We additionally compute unsatisfied states as described in \cref{sec:unsat_nor}.
    \item \textsc{Tiga(+Unsat)}: The \texttt{verifyta} engine distributed with \uppaal version 5.0.0~\cite{Behrmann2007Tiga,Hendriks2006Uppaal} and specifically the \tiga algorithm.\footnote{\tiga always keeps track of losing states without a strategy, i.e. the \textsc{Unsat} modification.}
\end{itemize}

The benchmark suite consists of $3$ types of TMG models, each instantiated in various sizes and with multiple associated TATL properties, totaling 236 model and query executions. The models are:
\begin{itemize}
    \item Train Gate~\cite{Alur99TA}: $N$ trains share the same bridge and a controller signals the trains if the bridge is occupied. Each train and the controller is a different player. Example queries include:
    \begin{itemize}
        \item $\coal{T_1,T_2,T_3}\Diamond num\_crossing\geq 1$; Can train $T_1$, $T_2$, and $T_3$ create a situation where more than one train is crossing at the same time? (No)
        \item $\coal{T_1}\Diamond crossed(T_1)$; Can train $T_1$ ensure that it eventually crosses the bridge? (No)
    \end{itemize}
    \item Mexican Standoff~\cite{Carlsen2023CGAAL} extended with reload time: $N$ cowboys are shooting at each other and must reload their gun between each shot. Each cowboy is a different player and the gun reloading system is also a player. Example queries include:
    \begin{itemize}
        \item $\coal{C_1}(alive(C_1)\,\until\, t > 1)$; Can cowboy $C_1$ guarantee staying alive for 1 second? (No)
        \item $\coal{Guns, C_1,\dotsc, C_{\lceil N/2\rceil}}\Box\neg\bigvee_{c\in \{C_1,\dotsc, C_{\lceil N/2\rceil}\}}alive(c)$; If half the cowboys cooperate and the gun reloading is on their side, can they ensure one of them survives? (Yes)
    \end{itemize}
    \item Phase King~\cite{Berman1993PhaseKing}: A consensus algorithm for $N$ nodes where time is divided into phases and rounds, and the king of each phase decides the voting tiebreaker. Each node is controlled by a unique player. Example queries include:
    \begin{itemize}
        \item $\coalbr{\;}\Diamond consensus(n_1,n_2,n_3)$; Node $n_1$, $n_2$, $n_3$ will reach consensus in some trace? (Yes)
        \item $\coal{}\Diamond consensus(n_1,n_2,n_3)$; Node $n_1$, $n_2$, $n_3$ will reach consensus in all traces? (No)
        \item $\coal{}\Box(\neg consensus(n_1, n_2, n_3) \lor \coal{n_1,n_2,n_3}\Box consensus(n_1, n_2, n_3))$; If nodes $n_1$, $n_2$, and $n_3$ have consensus, can they can stay in consensus? (Yes, if $N\leq5$)
    \end{itemize}
\end{itemize}

\paragraph*{Evaluation}

\begin{figure}[t]
    \centering
    \begin{subfigure}{0.5\textwidth}
        \centering
        \includegraphics[width=\textwidth]{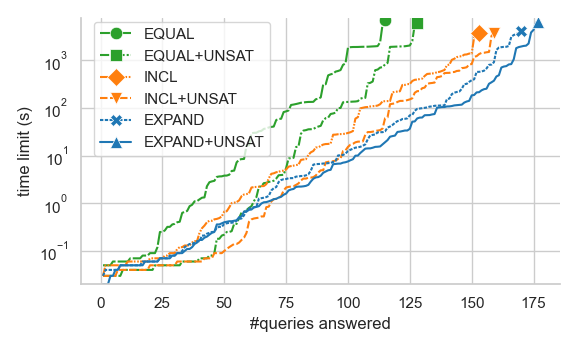}
        \caption{Time usage}
    \end{subfigure}%
    \begin{subfigure}{0.5\textwidth}
        \centering
        \includegraphics[width=\textwidth]{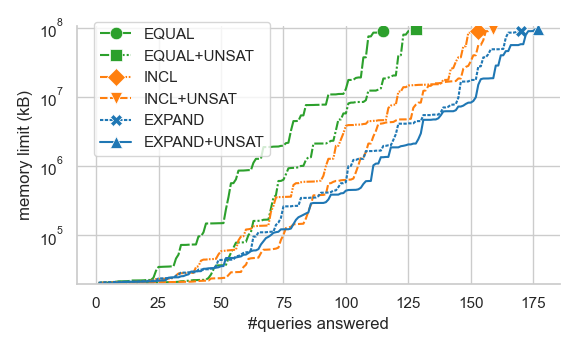}
        \caption{Memory usage}
    \end{subfigure}
    \caption{Cactus plots showing how many queries are answered within the indicated per-query resource limit. A configuration performs better if it can answer more queries using fewer resources.}
    \label{fig:cactus}
\end{figure}

\begin{figure}[t]
    \centering
    \begin{subfigure}{0.5\textwidth}
        \centering
        \includegraphics[width=\textwidth]{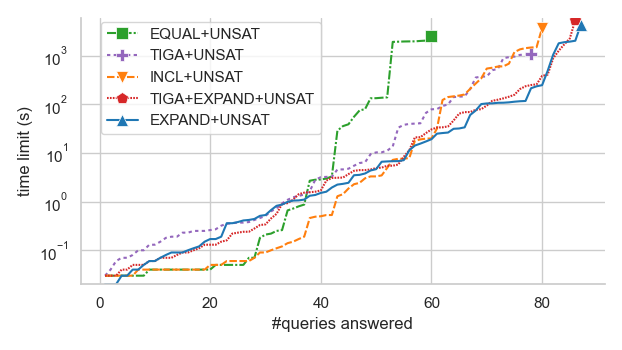}
        \caption{Time usage}
    \end{subfigure}%
    \begin{subfigure}{0.5\textwidth}
        \centering
        \includegraphics[width=\textwidth]{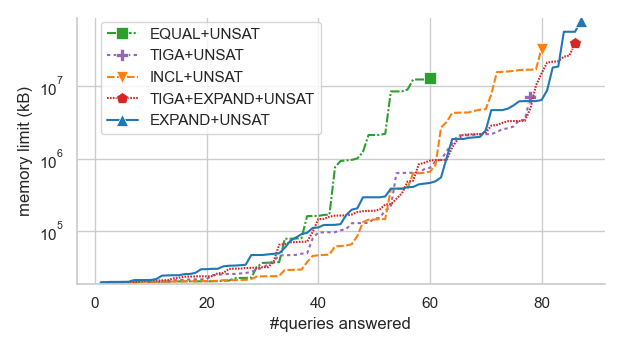}
        \caption{Memory usage}
    \end{subfigure}
    \caption{Cactus plots showing how many \tiga-compatible queries are answered within the indicated per-query resource limit. A configuration performs better if it can answer more queries using fewer resources.}
    \label{fig:cactus_tiga_queries}
\end{figure}

Performance comparison of the configurations can be seen in \cref{fig:cactus}. Here the query executions are ordered (independently for each method) by their running time. Note that the y-axis is logarithmic. The \textsc{Incl} configuration using vertex merging is more than one order of magnitude faster than the \textsc{Equal} configuration using no merging, and \textsc{Expand} is almost another order of magnitude faster for the most difficult instances. The \textsc{Unsat} modification further improves the three configurations matching the conclusion in~\cite{Cassez2005TimedGames}. We find a similar improvements for memory usage.
As described in \cref{sec:expand_abstraction}, \textsc{Expand} may explore edges unnecessarily, since the abstraction enables all transitions in a location, assuming they will be relevant eventually. In the plots, we see that the \textsc{Expand} configuration performs worse on easy queries since these extra edges hinder early termination. However, after just 1 second this downside is eliminated by the advantages of the method.

Of our 236 total queries, 150 have no nested coalitions and can be solved by \uppaaltiga as well. However, for a large portion of them, modifications to the TMG model are required to get the edges' controllability to correspond to the actions of the coalition. In \cref{fig:cactus_tiga_queries}, we compare the performance with \tiga on this subset of 150 queries. As expected, \textsc{Incl+Unsat} and \textsc{Tiga+Unsat} have similar performance on challenging queries, as these two configurations are very similar in practice. Our implementation seems slightly faster on easy queries. When we add the \textsc{Expand} abstraction to \tiga and remove its now redundant inclusion checking (configuration \textsc{Tiga+Expand+Unsat}), the runtime performance matches that of our \textsc{Expand+Unsat}, almost an order of magnitude faster than the previous state-of-the-art \tiga implementation.

A reproducibility package is available at~\cite{Jensen2025ReprodPack}.
\section{Conclusion}

We presented an encoding of the timed alternating-timed temporal logic problem in the extended abstract dependency graph (EADG) framework. This involved combining previous work on timed games, timed CTL, and alternating-time temporal logic.
Our work is thus an example of how various encodings of model-checking problems in the EADG framework are orthogonal and can be combined to solve the combined logic extensions. We also took this opportunity to provide many details left out in the previous paper on \uppaaltiga\cite{Cassez2005TimedGames}.
Furthermore, we formalized a generalization of conventional inclusion checking for the EADG framework. The resulting vertex merging technique can be used to remove vertices that can be derived from other vertices, which is especially useful when the derivation is cheaper than the computation of the value function of the removed vertex. Other domains where we foresee vertex merging being useful include model checking for Petri nets where one marking can cover other markings with less behavior.
The vertex merging also allowed us to easily show that we can better exploit the symbolic representation of states using an expansion abstraction.
In essence, our abstraction simplifies our symbolic states to discrete locations and their invariants.
Hence, \emph{all} valuations satisfying the given property in the location are propagated backward through the dependency graph, instead of restricting it to valuations that the exploration currently considers reachable. Thus, this abstraction makes the algorithm slightly closer to a traditional backward algorithm.

Our implementation and experiments showed that inclusion checking improves the performance of the naive encoding, while the expansion abstraction outperforms it by almost an additional order of magnitude. We also found that our algorithm is comparable to state-of-the-art \uppaaltiga when both use inclusion checking. By integrating our expansion abstraction in \tiga we also improved its performance by almost an order of magnitude.
The algorithms presented in this paper will be made available in an upcoming release of \uppaal.

\bibliography{bib}
\end{document}